\documentclass[]{spie}  %>>> use for US letter paper
\usepackage[]{graphicx}
\usepackage{epstopdf}
\usepackage{multirow}
\usepackage{rotating}

\title{Technology advancement of the CCD201-20
EMCCD for the WFIRST coronagraph instrument:
sensor characterization and radiation damage}

\author{Leon K. Harding\supit{1,a}, Richard T. Demers\supit{1,a}, Michael Hoenk\supit{a}, Pavani Peddada\supit{a}, Bijan Nemati\supit{a}, Michael Cherng\supit{a}, Darren Michaels\supit{a}, Leo S. Neat\supit{a}, Anthony Loc\supit{a}, Nathan Bush\supit{b}, David Hall\supit{b}, Neil Murray\supit{b}, Jason Gow\supit{b},  Ross Burgon\supit{b}, Andrew Holland\supit{b}, Alice Reinheimer\supit{c}, Paul R. Jorden\supit{d} and Douglas Jordan\supit{d}
\skiplinehalf
\supit{a}Jet Propulsion Laboratory, California Institute of Technology, 4800 Oak Grove Drive, Pasadena 91109, CA, USA; \\
\supit{b}Center for Electronic Imaging, Department of Physical Sciences, The Open University, Walton Hall, Milton Keynes, MK7 6AA, UK; \\
\supit{c}e2v inc., 765 Sycamore Drive, Milpitas CA 95035, USA; \\
\supit{d}e2v technologies, 106 Waterhouse Lane, Chelmsford, Essex, CM1 2QU, England
}

\authorinfo{$^{1}$Corresponding authors: Leon.K.Harding@jpl.nasa.gov; Richard.T.Demers@jpl.nasa.gov}
\begin{document} 
  \maketitle 

%%%%%%%%%%%%%%%%%%%%%%%%%%%%%%%%%%%%%%%%%%%%%%%%%%%%%%%%%%%%% 
\begin{abstract}
The Wide Field InfraRed Survey Telescope-Astrophysics Focused Telescope Asset (WFIRST-AFTA)
mission is a 2.4-m class space telescope that will be used across a swath of astrophysical research domains.
JPL will provide a high-contrast imaging coronagraph instrument—one of two major astronomical instruments. In
order to achieve the low noise performance required to detect planets under extremely low flux conditions,
the electron multiplying charge-coupled device (EMCCD) has been baselined for both of the coronagraph's
sensors—the imaging camera and integral field spectrograph. JPL has established an EMCCD test laboratory
in order to advance EMCCD maturity to technology readiness level-6. This plan incorporates full sensor characterization,
including read noise, dark current, and clock induced charge. In addition, by considering the unique
challenges of the WFIRST space environment, degradation to the sensor's charge transfer efficiency will be
assessed, as a result of damage from high-energy particles such as protons, electrons, and cosmic rays.
Science-grade CCD201-20 EMCCDs have been irradiated to a proton fluence that reflects the projected
WFIRST orbit. Performance degradation due to radiation displacement damage is reported, which is the first
such study for a CCD201-20 that replicates the WFIRST conditions. In addition, techniques intended to identify
and mitigate radiation-induced electron trapping, such as trap pumping, custom clocking, and thermal cycling,
are discussed.

\end{abstract}

%>>>> Include a list of keywords after the abstract 

\keywords{electron multiplying charge-coupled devices; WFIRST-AFTA; radiation damage; L2 orbit; techniques: mitigation\\\\ \textit{Published in J.
Astron. Telesc. Instrum. Syst., 2, 1, 2016}}

%%%%%%%%%%%%%%%%%%%%%%%%%%%%%%%%%%%%%%%%%%%%%%%%%%%%%%%%%%%%%
\section{Introduction}
\label{sec:intro}  % \label{} allows reference to this section

\subsection{Wide Field InfraRed Survey Telescope--Astrophysics Focused Telescope Asset
Coronagraph}

The Wide Field InfraRed Survey Telescope--Astrophysics
Focused Telescope Asset (WFIRST-AFTA) is a NASA space
observatory that has been designed to probe dark energy, to
carry out wide-field near infrared (NIR) surveys, and to discover
and characterize extrasolar planets (hereafter exoplanets) in the
visible spectrum. WFIRST-AFTA will make use of an existing
2.4-m aperture telescope and will advance what is currently possible
in exoplanet spectral characterization and imaging. The
telescope feeds an instrument suite consisting of a wide-field
imager (WFI) and a coronagraph instrument (CGI). These two
instruments are complementary. The WFI will collect NIR statistical
data on planetary systems over large regions of the sky
using gravitational microlensing, whereas the coronagraph will
carry out direct imaging and detailed visible spectroscopy of
a sample of exoplanets. The coronagraph will be able to detect
and characterize cold Jupiters, mini-Neptunes, and possibly super-Earths\cite{spergel15} and for the first time directly image planets analogous
to those in our solar system. This spectroscopic characterization
will reveal the atmospheric composition of these planets
and will be used to search for spectral signatures of life. In addition,
the coronagraph will be used to characterize debris disks in
and around planetary orbits—an important tool that can be used
to improve our understanding of planet formation.

In order to adequately form images and collect spectra of
exoplanets, it is necessary to suppress the starlight to a contrast
of order 10$^{-9}$ for cold Jupiters, mini-Neptunes, and super-
Earths. The coronagraph accomplishes this by using (1) a series
of masks to reject starlight, to block diffracted light, and to
reduce starlight speckles, and (2) an adaptive optics system,
employing a pair of deformable mirrors (DMs) in order to eliminate
residual speckles due to optical imperfections in the entire
optical beamtrain. The result is a high-contrast point-spread
function (PSF) with a dark hole between inner and outer working
angles allowing observations of faint companions.

The WFIRST-CGI can operate in two modes: (1) a hybrid
Lyot coronagraph (HLC) for exoplanet photometry and discovery,
and (2) a shaped pupil coronagraph (SPC) for exoplanet
spectroscopy and debris disk characterization. The HLC uses
a focal plane phase-amplitude mask, whereas the SPC uses a pupil mask with binary apodization. The flight instrument comprises
a single optical beam train that will be operated in either
the HLC or SPC mode. The CGI switches between the two
operational configurations using active rotary mechanisms to
change pupil masks, focal plane masks, Lyot masks, bandpass
filters, and an actuated fold mirror to select either the imaging
camera or the integral field spectrograph (IFS). In both the
HLC and SPC modes, the pair of DMs is used for closed-loop
reduction of speckle. This dual mode architecture enables the
optimum coronagraph configuration to be used for the different
types of planet/disk characterization. The two coronagraph
architectures are complementary. The HLC can achieve deeper
contrast at certain working angles, whereas the SPC is relatively
insensitive to jitter and has lower chromaticity.

\begin{figure}[!t]
   \centering
   \includegraphics[width=14cm]{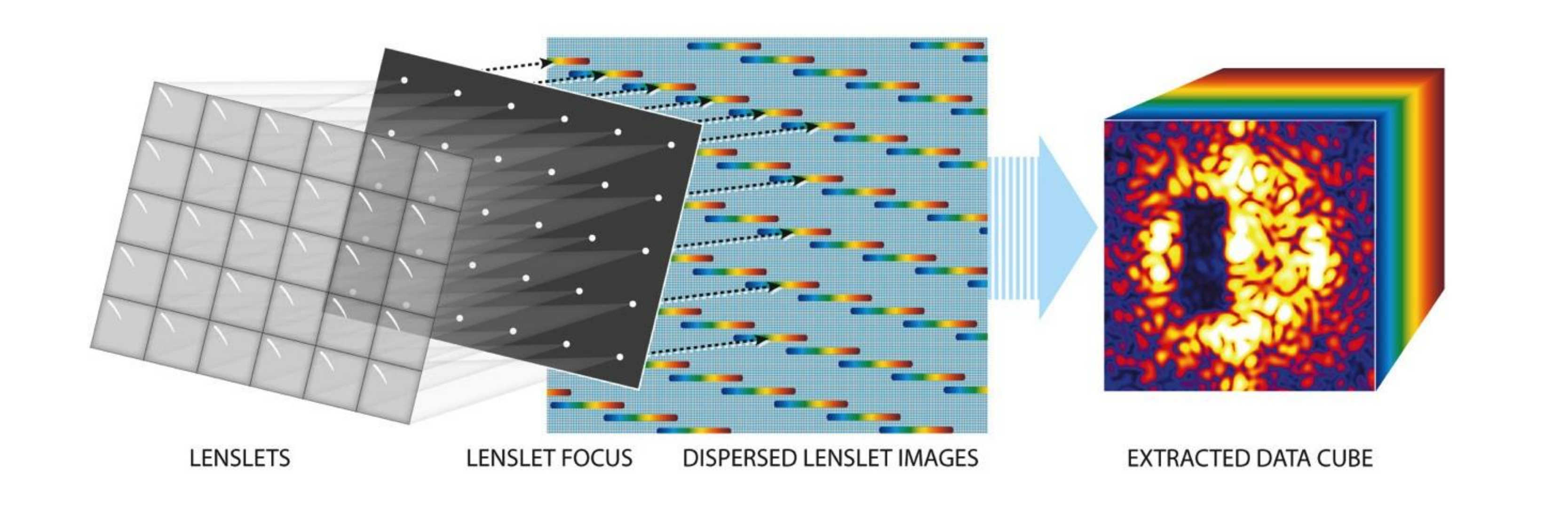}
\caption{A lenslet array with a pinhole mask is clocked with respect to the rows of the electron multiplying
charge-coupled device (EMCCD) in order to optimize the separation between neighboring spectra on
the detector thereby reducing spectral crosstalk. A pinhole mask at the exit surface of the lenslet array
serves to truncate the tails of the lenslet point-spread functions in order to reduce spectral crosstalk on
the detector. A spectral data cube is extracted in postprocessing. Figure provided by Michael McElwain of
Goddard Space Flight Center (GSFC).}\label{fig:AFTA-lenslet}
\end{figure}

In order to maintain the contrast stability over observation
times, a low-order wavefront sensor and control system in
the coronagraph is used to sense and correct lower-order wavefront
error resulting from (1) thermally-induced optical misalignment
and optical surface distortions, and (2) line-of-sight
(LOS) jitter caused by vibration sources, such as the spacecraft
reaction wheels and the WFI cryocooler. A Zernike wavefront
sensor uses the rejected starlight to sense aberrations from Z1,2
(x- and y-tilts) up to and including Z11 (spherical aberration).
The x- and y-tilts resulting from LOS jitter are corrected using
a fast steering mirror, whereas the low-order wavefront drifts are
corrected by a pistoning fold mirror for focus correction and
the DMs for other low-order aberrations, such as astigmatisms
and comas.

When carrying out planet discovery, photometry, and disk
imaging, the HLC or SPC relays an image of the high-contrast
“dark hole” onto the imaging camera. A stationary polarizing
beam splitter in front of the imaging camera separates the
dark hole image into two orthogonal polarizations, whose two
resulting images fill a small rectangular area of the detector.
When taking planetary spectra, by contrast, the IFS images a
regular grid of spectrally dispersed PSFs onto the detector.
Several aspects of the IFS design\cite{gong15} further reduce the already
very low-light level of the planet image from the coronagraph.
The IFS separately collects spectra across the visible spectrum in
three separately detected 18\% wide sub-bands centered at 660,
770, and 890 nm. At the input of the IFS is a lenslet array used
for spatial sampling the dark hole, where the PSF sampling on
the lenslet array is 3 $\times$ 3 lenslets per λ∕D (PSF core). A two-element
prism spectrally disperses the PSFs along a single axis, further spreading the light on the detector. Thus, the very low
expected photon count requires subelectron read noise, dark
current, and clock induced charge (CIC) to carry out planet
spectroscopy. The formation of the grid of spectra on the detector
is illustrated in Fig. 1.

\subsection{WFIRST-CGI Detector Selection}\label{sec:trade-study}

\subsubsection{Detector candidates}

Many astronomical instruments have used conventional charge coupled
devices (CCDs), segmented CCDs, or hybrid complementary
metal oxide semiconductor (CMOS) focal plane
arrays\cite{dhillon07,ives08,law06,matt10,odonoghue95,stover87,wilson03,rauscher07}. Conventional CCDs and hybrid VIS/IR CMOS\cite{beletic08,blank12}
devices have been the preferred focal plane arrays for both
ground-based and space-based telescopes and visible/IR photometric
instruments. Conventional CCDs have the advantage of
being generally easier to characterize due to their extremely
simple design. Older CCDs have less output taps and since
noise increases with bandwidth, older CCDs have higher read
noise. More recent CCDs have more output taps and are therefore
less noisy and their primary source of noise is the output
driver. The charge to voltage conversion in conventional CCDs
occurs on the gate capacitor of a metal oxide semiconductor
field-effect transistor (MOSFET) or junction gate field-effect
transistor (JFET), which is inherently nonlinear, but if running
high-power supply voltages with low-voltage signal swings,
CCDs have linear response. In the design of CCDs, conversion
gain, output driver strength, and bandwidth are linked and
so design optimization is limited. Hybrid CMOS detectors,
owing to their complex design, can be difficult to characterize
but do not suffer from noise due to either charge transfer inefficiency
(CTI) or CIC. The breed of CMOS sensors used in
astronomy--typically source follower pixels--has high gain
in the pixel and is therefore generally limited by pixel noise
and insensitive to read noise in the column buffer. Source
follower CMOS pixels typically use a gate capacitor with a
MOSFET to convert charge to voltage and like CCDs are inherently
nonlinear. Their response, however, is linear for most of
the electron well range with nonlinearity for nearly empty and
nearly full wells. In CMOS sensor design, the conversion gain,
output driver strength, and bandwidth can be separately traded
to optimize performance.

With the advent of the electron multiplying CCD (EMCCD)
at the beginning of the last decade\cite{jerram01,hynecek01} high signal to noise (S/N) performance became possible over much shorter timescales,
resulting in negligible effective read noise and low spurious
noise contribution. This performance is achieved by means
of a multiplication register that lies after the conventional serial
register where charge can be amplified by factors of thousands
via avalanche multiplication. We explain this phenomenon in
more detail in Sec. 2. In this sense, EMCCDs can outperform
conventional CCDs and are technologically mature since they
are produced in medium volume as a commercial product for
medical devices and for ground-based observatory instruments.
Nonetheless, EMCCDs have little or no space flight heritage.

In order to fully assess the S/N performance of the EMCCD
and to compare with other candidate sensors having reasonably
good performance but which may also have higher flight
heritage and/or higher technological maturity, we carried out
a trade study to compare the principal low-light level detector
technologies. Following a survey of such detectors, candidate
technologies were selected for this study. These candidates
were evaluated and ranked based on (1) compliance with
performance requirements for the Imager and IFS detectors,
(2) flight heritage, (3) desired format and pixel pitch, and
(4) planet yield predicted by a coronagraph model. The candidate
detectors are [in order of descending technology readiness
level (TRL)] scientific CCDs [Euclid, Hubble Space Telescope
(HST) wide-field camera 3 (WFC3) and Gaia], Hybrid Visible
Silicon Imager (JMAPS), EMCCDs, and scientific CMOS
detectors. A brief description of each of these technologies is
provided in this section and summarized in Table 1.\\

\begin{table}[!t]
\caption{Detector candidate technologies for JPL trade study. The read noise performances quoted are nominal values from respective studies
that were released. The ``Limitations'' column refers to either a nonoptimal design detail for the WFIRST-CGI or to some test or study that has not
yet been carried out. We note that at the time of the trade study, the CCD201-20 had not been radiation tested.}
\begin{center}
\begin{tabular}{cccccccc}
\hline
Short & Mission &Detector  & Manuf.  & TRL & Pixel & Full & Limitations \\
 Name & &Model/N & & & &Format& \\
  & & & & & ($\mu$m)& (pixel)& \\
\hline \hline \\
sCMOS&Ground-based& CIS2051& Fairchild  & 4 & 6.5 $\times$ 6.5 & 2560 $\times$ 2560 & Lowest TRL\\
&&&Imaging&&&& No rad./therm.\\
&&&&&&& cycling \\
&&&&&&& No temp data\\
\hline\\
Euclid CCD&Euclid vis.& CCD273-84& e2v  & 6 & 12 $\times$ 12 & 4096 $\times$ 4096 & RN $\sim$3.5 e$^{-}$ \\
&instrument&&&&&&\\
\hline\\
HST CCD&HST WFC3& CCD43& e2v  & 9 & 15 $\times$ 15 & 2050 $\times$ 4096 & RN $\sim$3 e$^{-}$ \\
&&&&&&&Std epi depth\\
\hline\\
Gaia CCD&Gaia& CCD91-72& e2v  & 9 & 30 $\times$ 10 & 1966 $\times$ 4500 & RN $\sim$6 e$^{-}$ \\
&&&&&&&Unfavorable\\
&&&&&&&format\\
\hline\\
JMAPS&JMAPS& HiViSI& Teledyne  & 6 & 18 $\times$ 18 & 1024 $\times$ 2048 & RN $\sim$5 e$^{-}$ \\
HyViSI&&&&&&&\\
\hline\\
EMCCD&WFIRST-CGI& CCD201-20& e2v  & 5 & 13 $\times$ 13 & 1024 $\times$ 2048 & Low TRL \\
Std AN&&&&&&&No therm. \\
&&&&&&&cycling\\
EMCCD&WFIRST-CGI& CCD201-20& e2v  & 2 & 13 $\times$ 13 & 1024 $\times$ 2048 & Thin epi (std) \\
DD AN&(DD)&&&&&&\\
\hline\\
\multicolumn{8}{c}{Acronyms: TRL=Technology Readiness Level; RN=Read Noise.}\\
\multicolumn{8}{c}{HST=Hubble Space Telescope; WFC3=Wide Field Camera 3; DD=Deep Depletion.}\\
\multicolumn{8}{c}{Abbreviations: Std=Standard; AN=Analog; Rad./Therm. =Radiation/Thermal; pix=pixel.}\\
\hline
\end{tabular}
\end{center}
\label{table:trade-study1}
\end{table}

\noindent \textbf{i. Scientific CCDs:}  Scientific CCDs offer the highest TRL,
having demonstrated excellent performance in multiple
missions, including the HST wide-field and planetary
camera 2, and WFC3 instruments, the Kepler focal
plane, and the Gaia suite of instruments. A key advantage
of scientific CCDs is the availability of a mature
deep depletion (DD) technology, which is based on
the use of high-purity silicon in the detector to enable
high quantum efficiency (QE) in the 800- and 900-nm
WFIRST-CGI bands. An important limitation of conventional
scientific CCDs is the relatively high read
noise, which is typically $>$2 e$^{-}$ pix$^{-1}$.\\
  
\noindent \textbf{ii. Hybrid Visible Silicon Imager (HyViSI$^{\small \mbox{TM}}$):} Unlike
CCDs, which require thousands of on-chip charge transfers
to form an image based on serial readout by one or
a few output amplifiers, CMOS detectors comprise
two-dimensional pixel arrays with a photodiode and
readout circuitry in each pixel. Teledyne imaging sensors
have developed a hybrid silicon CMOS detector
(the HyViSI™ detector\cite{bai12}) by interconnecting a visible
silicon photodiode array to a CMOS readout integrated
circuit (ROIC). HyViSI hybrids combine the high
responsivity of silicon photodiodes with CMOS parallel
readout architecture, which provides unique capabilities.
The correlated double sampling (CDS) read noise of
HyViSI is comparable to that of scientific CCDs. The
nondestructive parallel read capability of the CMOS
ROIC enables reduction of the standard CDS read noise
by recording multiple reads during integration (a.k.a.
Fowler sampling or sampling up the ramp, SUTR) and
fitting the results to a linear ramp. Such techniques
are widely used in astronomy to reduce the contribution
of reset and read noise. In addition, nondestructive
read capability enables easy and accurate removal of
cosmic rays (CRs) while preserving signal. Finally,
since HyViSI detectors do not transfer charge through
many pixels, they are not susceptible to deferred charge,
which is one of the main limitations of scientific CCDs
exposed to high radiation environments. HyViSI detectors
were hybridized to the H4RG-10 ROIC for the
U.S. Navy under the Joint Milli-Arcsecond Pathfinder
Survey (JMAPS) program\cite{gaume09}. Although the JMAPS mission
was canceled before launch, the H4RG-10 HyViSI
flight detectors attained TRL-6 status.\\

\begin{table}
\caption{Detector candidate performance parameters for JPL trade study. We note that the excess noise factor, or ENF, is an added variance
based on the EM process--this is explained further in Sec. 2. The read noise performances quoted are nominal values from commercial cameras,
instrument reports, or measurements at JPL. We note that CIC can differ greatly for these devices based on their mode of operation as well as other
factors. We measured nominal CIC for the CCD201-20 at 3 $\times$ 10$^{-3}$ e$^{-}$ pix$^{-1}$ fr$^{-1}$, and subsequently set all other applicable devices to this level as the best case for each. Dark current values stated correspond to nominal operating temperatures that are indicated in parentheses.}
\begin{center}
\begin{tabular}{cccccccc}
\hline
Performance & sCMOS  & Euclid CCD  & HST CCD & Gaia CCD & JMAPS&EMCCD&EMCCD \\
Parameter &  & &  &  &HiViSI&Std AN & DD AN \\
\hline \hline \\
ENF&1&1&1&1&1&1.41&1.41\\
\hline\\
CIC&0 &3 $\times$ 10$^{-3}$&3 $\times$ 10$^{-3}$&3 $\times$ 10$^{-3}$&0&3 $\times$ 10$^{-3}$&3 $\times$ 10$^{-3}$\\
(e$^{-}$/pix/fr)&&&&&&&\\
\hline\\
I$_{dk}$&7 $\times$ 10$^{-3}$ &5.5 $\times$ 10$^{-4}$&5.5 $\times$ 10$^{-4}$&1 $\times$ 10$^{-3}$&1 $\times$ 10$^{-3}$&3 $\times$ 10$^{-5}$&5 $\times$ 10$^{-4}$\\
(e$^{-}$/pix/sec)&(233 K)&(153 K)&(190 K)&(158 K)&(170 K)&(165 K)&(165 K)\\
\hline\\
RN&1.5&3.6&3.1&4.6&5&0.2$^{\dag}$&0.2$^{\dag}$\\
(e$^{-}$)&&&&&&&\\
\hline\\
QE, 550 nm & 55 & 83 & 63 & 90 & 84& 92&92\\
(\%) &&&&&&&\\
\hline\\
QE, 660 nm & 52 & 83 & 65 & 89 & 89& 90&90\\
(\%) &&&&&&&\\
\hline\\
QE, 890 nm & 12 & 40 & 25 & 24 & 85& 27&53\\
(\%) &&&&&&&\\
\hline\\
\multicolumn{8}{c}{$^{\dag}$Using Electron Multiplication (EM) gain of $\sim$450 based on an output amplifier}\\
\multicolumn{8}{c}{read noise of 89.78 e$^{-}$ at a 10 MHz read out rate (see Table~\ref{table:RN}).}\\
\hline\\
\end{tabular}
\end{center}
\label{table:trade-study2}
\end{table}

\begin{figure}[t!]
   \centering
   \includegraphics[width=10.9cm]{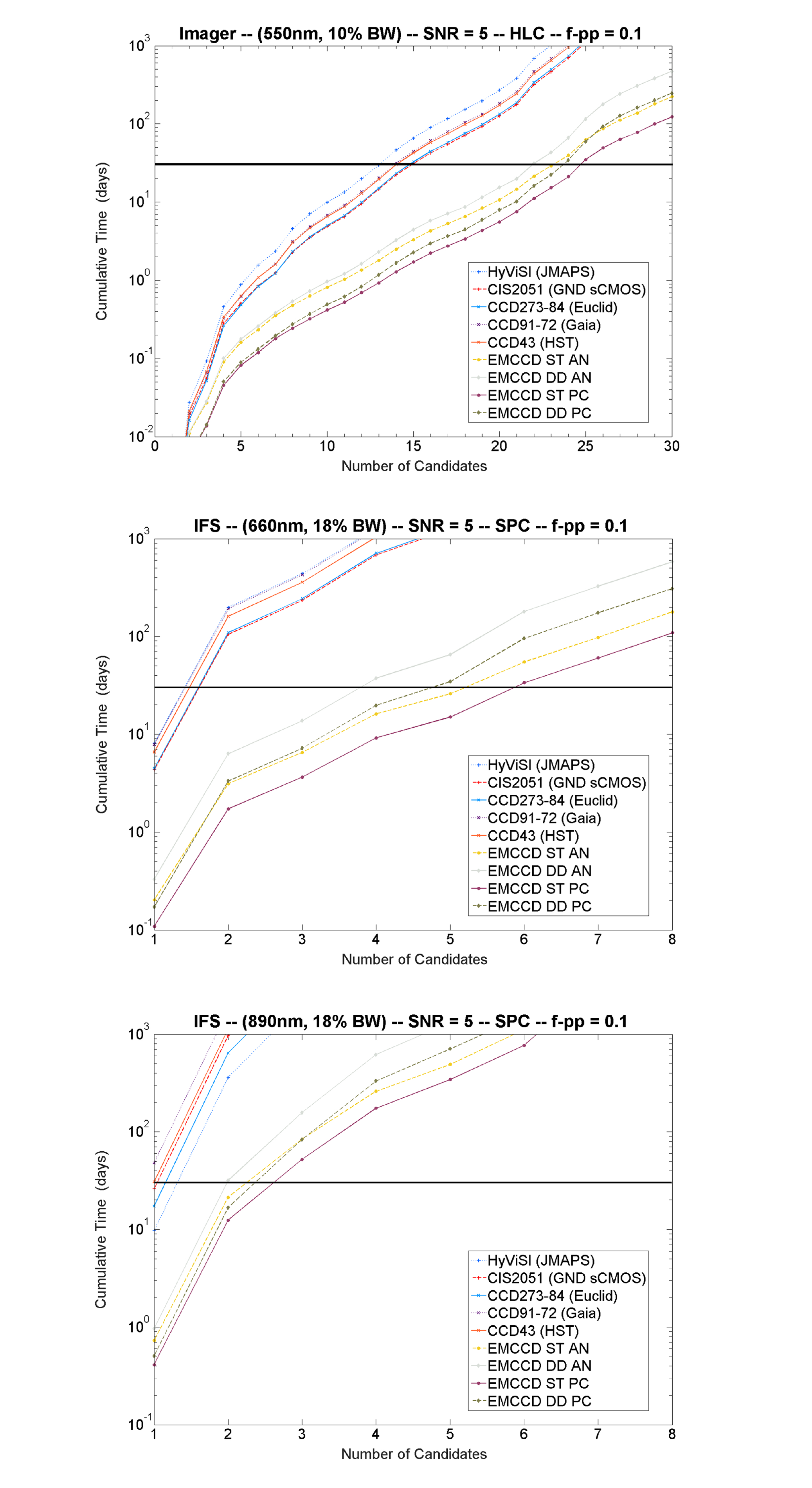}
\caption{Analytical prediction of observation time required to achieve
S/N $=$ 5 versus the number of planets for each of the detector candidates.
(a) 550-nm 10\% Imager band (HLC), (b) 660-nm IFS 18\%
band (SPC), and (c) 890-nm IFS 18\% band (SPC). f-pp denotes the
postprocessing residual error and is assumed to have a value of 0.1
consistent with contrast improvement (reduction) of a factor of 10.
The horizontal black line indicates a 30-day threshold.}\label{fig:trade-study-plots}
\end{figure}

\noindent \textbf{iii. Electron Multiplying CCDs (EMCCDs):} EMCCDs are a variant of scientific CCDs, in
which a specially designed serial output register provides
amplification using avalanche multiplication of
collected charge. Charge amplification in EMCCDs
significantly improves low-light level sensitivity by
effectively suppressing read noise by many orders of
magnitude [depending on the electron multiplication
(EM) gain]. Although EMCCDs are less mature than
scientific CCDs, e2v manufactures EMCCDs using the same materials, designs, and processes developed for
high-heritage scientific CCDs. Some devices contain
important design variations relevant to the space environment;
however, the CCD201-20 (see Table 1) is
not made with e2v's radiation-hard oxide process
(see Sec. 4.2.1 for further discussion). EMCCDs have
undergone extensive development and testing for
ground-based astronomical instruments and medical
applications. One EMCCD has been tested in radiation
environments\cite{michaelis13} comparable to the planned WFIRST
orbit. e2v can provide customized DD EMCCDs, enabling
high QE in the 800- to 980-nm band. However,
DD EMCCDs are less mature than DD conventional scientific
CCDs. Unlike the latter, DD EMCCDs require a
special front-side implant. Additionally, DD CCDs and
EMCCDs are incompatible with bias voltages required
for inverted mode operation (IMO). Consequently, DD
EMCCDs must be operated in noninverted mode operation
(NIMO), which results in far higher surface dark
current, the dominant contributor to dark current. \\
  
\noindent \textbf{iv. Monolithic CMOS detectors:} Until relatively recently,
monolithic visible CMOS detectors were too noisy to
compete with CCDs in low-light level imaging applications.
However, advances in scientific CMOS detector
design and fabrication led to the development of
CMOS detectors with subelectron read noise. In 2009,
Fairchild Imaging teamed with Andor Technology and
PCO AG to develop a commercially available line of
scientific monolithic CMOS detectors and cameras that
are competitive with EMCCDs in low-light level imaging
applications. Monolithic CMOS detectors have all
of the advantages offered by parallel read architecture,
as described nearby for the HyViSI CMOS detectors.\\

%%%%%%%%%%%%%%%%%%%%
\begin{sidewaystable}
\caption{Detector trade matrix showing 12 discriminator categories, labeled a to l. Each option was separately scored against each discriminator.
The scores were categorized as follows: Best (best performance), Wash (indistinguishable from Best), Small (lower performance but small difference from Best), Sig (lower performance but significant difference from Best), and Very lg (lower performance but very large difference from Best). These rankings were assigned numerical scores (Best $=$ 4, Wash $=$ 4, Small $=$ 3, Sig $=$ 2, Very lg $=$ 1).}
\begin{center}
\begin{tabular}{cccccccccc}
\hline\\
\multicolumn{10}{c}{\textbf{Decision Statement: Select a detector model for the flight IFS camera.}}\\
\hline\\
&&&Option 1& Option 2& Option 3& Option 4& Option 5& Option 6 & Option 7\\
\hline\\
ID& Discriminator&Weights&sCMOS&Euclid CCD&HST CCD&Gaia CCD & JMAPS &EMCCD AN&EMCCD\\
&&(\%)&&&(WFC3)&&HiViSI&&DD AN\\
\hline \hline\\
a&RN$<$0.2 e$^{-}$/pix&6&Sig&Very lg&Very lg&Very lg&Very lg&Wash&Best\\
\hline\\
b&CIC$<3\times 10^{-3}$ e$^{-}$/pix/fr&5&Best&Wash&Wash&Wash&Wash&Wash&Wash\\
\hline\\
c&I$_{dk}<5\times10^{-4}$ e$^{-}$/pix/sec & 5 & Sig & Sig & Sig& Very lg &Sig &Best&Sig\\
\hline\\
d& QE, 550 nm & 5 & Very lg &Small &Very lg& Wash &Very lg& Wash& Best\\
\hline\\
e& QE, 660 nm & 5 & Very lg &Small &Very lg& Sig &Best& Sig &Wash\\
\hline\\
f& QE, 890 nm & 5 & Very lg &Very lg &Very lg& Very lg &Best& Very lg & Very lg\\
\hline\\
g& Reach TRL 6 (July 2016) & 20 & Sig &Best &Wash& Wash &Wash &Sig& Very lg\\
\hline\\
h& Format: 1024 $\times$ 1024 (min)&3& Wash &Wash &Wash &Wash &Wash& Wash& Best\\
\hline\\
i& Pixel pitch (sq): 13 -- 15 $\mu$m & 3 & Sig& Small& Wash &Sig& Small &Best &Wash\\
\hline\\
j&\# planets imaged (30 dys)&15&Very lg & Very lg & Very lg& Very lg&Very lg&Best&Wash\\
&(Imager, 550 nm)&&&&&&&&\\
\hline\\
k&\# planets spectrally char.&14&Very lg & Very lg & Very lg& Very lg&Very lg&Best&Sig\\
&(30 dys, IFS, 660 nm)&&&&&&&&\\
\hline\\
l&\# planets spectrally char.&14&Small & Small & Small& Small&Small&Best&Wash\\
&(30 dys, IFS, 890 nm)&&&&&&&&\\
\hline\\
&\textbf{Weighted Sum}&\textbf{100}&\textbf{1.9}&\textbf{2.4}&\textbf{2.3}&\textbf{2.4}&\textbf{2.5}&\textbf{3.4}&\textbf{2.9}\\
\hline
\end{tabular}
\end{center}
\label{table:trade-study3}
\end{sidewaystable}

%%%%%%%%%%%%%%%%%%%%
\begin{table}
\caption{Evaluation of detector risks and opportunities.}
\begin{center}
\begin{tabular}{ccccccccc}
\hline\\
&&Option 1& Option 2& Option 3& Option 4& Option 5& Option 6 & Option 7\\
\hline\\
ID&Descrip. &sCMOS&Euclid &HST CCD&Gaia CCD & JMAPS &EMCCD&EMCCD\\
&&&CCD&(WFC3)&&HiViSI&AN&DD AN\\
\hline\\
\multicolumn{9}{c}{\textbf{LIKELIHOOD}}\\
\hline \hline\\
Risk 1& Low yield/ & Low & Low & Low & Low & Low & Low & Medium\\
&failed lot & & & & & & &\\
\hline\\
Risk 2&Low&Low & Low & Low & Low & Low & Low & Medium\\
&operability & & & & & & &\\
\hline\\
Risk 3&Radiation&Low& Medium &Medium &Medium &Medium &Medium &Medium\\
&degradation&&&&&&&\\
&(EOL)&&&&&&&\\
\hline\\
Oppor-&Science&Low &Low &Low &Medium &Medium&High &High\\
tunity 1&discovery&&&&&&&\\
\hline
\end{tabular}
\end{center}
\label{table:trade-study4}
\end{table}
%%%%%%%%%%%%%%%%%%%%

\subsubsection{Detector trade study}

The trade study was composed of seven specific detector models
from four competing technologies. The seven candidates were
as follows:\\

\noindent \textbf{1)}	Monolithic scientific CMOS. \\\\
\textbf{2)}	Conventional CCDs as used in EUCLID's visible focal plane array. \\\\
\textbf{3)}	Conventional CCDs as used in HST's WFC3. \\\\
\textbf{4)}	DD CCDs as used in Gaia's large Astrometric Focal Plane (AFP\cite{kohley12}). \\\\
\textbf{5)}	Hybrid Silicon CMOS arrays as used in the JMAPS focal plane array.\\\\
\textbf{6)}    Standard silicon-thickness EMCCD operated in analog EM gain mode and photon counting mode.\\\\
\textbf{7)}    DD EMCCD operated in analog EM gain mode and photon counting mode.\\

To evaluate the relative science yield of the candidate detectors,
we simulated end-to-end CGI performance using each of
seven specific detector candidates. The model predicted the time
required to collect images and spectra of planets from a list of
known radial velocity (RV) planets. The predicted observation
time was calculated as the time required to achieve a S∕N of
5 for either planet imaging or spectroscopy. The performance
parameters that were used as model inputs appear in Table 2.
All parameters were taken from available literature or from
measurements collected by the authors (Fairchild Imaging CIS2051 specifications sheet, Endicott et al. 2012\cite{endicott12}, Kimble et al. 2009\cite{kimble09}, Prod'homme et al. 2011\cite{prodhomme11}, Gaume et al. 2009\cite{gaume09}, this work). For ease of
analysis, all CCDs were assumed to have the same value for
CIC, though in reality CIC between different CCDs can vary
an order of magnitude depending on whether they are operated
in NIMO or IMO, as well as pixel size, clock voltages, and operating
speed. However, CIC measurements for many of the CCD
candidates did not exist because it is not the leading contributor
to overall S/N. Our assumption of the same CIC for all of the
CCD candidates represents a best case for each. We find in
the end that this assumption does not impact the final result.
In addition to the seven candidates listed previously, the model
included simulation of both the standard thickness and DD EMCCD as operated in photon-counting mode. Thus, a total of
nine different cases were analyzed using the model. The underlying
model assumptions are as follows: \\

\noindent \textbf{1)} HLC design version of January 2014; this HLC design
has similar contrast in working angles to the current
HLC design. The coronagraph focal plane mask
design in the 2014 HLC design was not easy to fabricate.
The current version mask design is easily fabricated.
This does not affect the detector comparison.\\ \\
\textbf{2)} SPC design version of December 2013. \\ \\ 
\textbf{3)} Exoplanet detection targets were taken from a known
list of RV planets sorted for the model in order of
increasing difficulty of detection, as measured by
the time required to achieve S/N of 5. \\ \\
\textbf{4)} The exoplanet zodiacal background is the same as the
local zodiacal background, i.e. 22 mag arcsec$^{-2}$.\\ \\
\textbf{5)} A factor of 10 improvements in contrast results from
data postprocessing. The postprocessing factor, f-pp,
is assumed to have a value of 0.1 consistent with
contrast improvement (reduction) of a factor of 10.
This is the largest single contributor to the model
uncertainty but this uncertainty affects all candidates
equally.\\

The analytical model predictions of planet yield are shown in
Fig. 2. The model showed that for the expected science targets,
the IFS detector will sense a mean signal on the order of
3 $\times$ 10$^{-3}$ e$^{-}$ pix$^{-1}$ sec$^{-1}$. The target planets would be imaged using
the imaging camera and would be spectrally characterized in
each of the three IFS bands in the presence of stellar zodiacal
background and detector noise. In Fig. 2, the model results are
shown separately for the imaging channel fed by the HLC and
for the two of the three IFS spectral bands that are fed by the
SPC. The imaging channel uses a 10\% spectral band centered at
550 nm, whereas the two analyzed IFS bands are centered at
660 and 890 nm, each with 18\% spectral bandwidth. It is seen
from the analytical results that among the seven trade study
candidates, both the standard thickness and DD EMCCD outperformed
all the other detector candidates in each of the
three observational cases, regardless of whether the EMCCD
was operated with analog gain or in photon-counting mode.
In each case, the best performer was the standard thickness
EMCCD in photon-counting mode. There is a significant performance
gap between the four EMCCD variants and the other
five detectors, whose performance was quite similar. For the
WFIRST-CGI, the EMCCD has the best overall performance.

For the selection of the baseline CGI flight detector, the
seven candidates were scored separately in 12 discriminator
categories: six device performance parameters, flight heritage
(TRL), pixel size, array format, and the science yield as measured
by number of planets detected in 30 days in each of
three channels. All of these discriminator scores were weighted
and summed for an overall score. The highest weighted discriminators
were (1) flight heritage (ability to reach TRL-6 by July
2016), weighted at 20\% and (2) the time required to image and
spectrally characterize planet targets, weighted at 43\%. The
highest scoring candidate was the DD EMCCD followed by
the standard thickness EMCCD. The scoring with weights
and summary scores are tabulated in Table 3. The EMCCD outperforms the other candidates because of its superior S/N
at CGI photon flux levels, largely due to the order of magnitude
reduction in read noise resulting from avalanche charge
multiplication. The standard thickness EMCCD scored highest, followed by the DD EMCCD. The next lowest scores correspond
to the three conventional CCDs and the HyViSI
CMOS detector. The scoring differences within this subgroup
are not deemed significant. The lowest score belongs to the
monolithic CMOS detector. Commercially available scientific
monolithic CMOS detectors have low QE in the 890-nm band.
This is not a fundamental limit of monolithic CMOS technology.
The QE of CMOS is fundamentally limited by the optical
properties of silicon and the thickness of the absorbing layer.

The seven detector candidates were also evaluated in relation
to risks and opportunities, as shown in Table 4. The standard
thickness EMCCD was assessed to have manageable risk with
high opportunity in science yield. The DD EMCCD offers one
performance advantage over the standard thickness version. As
shown in Table 2, the QE for the DD device is higher in the
770- and 890-nm bands. However, e2v has never fabricated
the CCD201-20 with DD silicon and therefore, selection of
this device would entail risks of low fabrication yield or low device operability. Additionally, a DD device cannot be run in
IMO; therefore, dark current contribution is significantly higher
and far outweighs the benefit of greater QE response in the NIR,
consistent with the predictions in Fig. 2. (We discuss this in
more detail in Secs. 5.3 and 5.5.3.) The standard thickness
version of CCD201-20 is a standard product fabricated in
medium volume for commercial applications and therefore has
well characterized yield and low fabrication process risk. As
a result of this evaluation process, the selection of the EMCCD
as the imaging and IFS flight detector stands on firm ground.

\subsection{Detector Technology Development Plan}

Under NASA prephase-A technology development funding,
JPL is moving forward the technological maturity of the
CCD201-20 for space flight in order to reach NASA TRL-6.
A three-part program to advance the device maturity is under
way, consisting of (1) radiation environment testing, (2) noise
performance optimization, and (3) thermal environment testing.\\

\subsubsection{Radiation environment testing}

\textbf{Phase I:} In the first of a two-phase radiation test, the radiation
hardness of a pair of CCD201-20 engineering grade devices at
ambient temperature was characterized after having been irradiated
by a single exposure displacement damage dose (DDD)
representative of a 6-year mission with direct insertion to L2
orbit. The result is also relevant to a potential geosynchronous
orbit, as explained in Sec. 3. This first phase was carried out to
provide an early indication of survivability of the CCD201-20 in
the L2 radiation environment since no EMCCD has yet been
qualified or flown in space and, as mentioned previously, a
large format EMCCD of this design has not yet been radiation
tested. The results of phase I are relevant to a mission in which
periodic warm cycling of the coronagraph EMCCDs would
be conducted to reverse some of the radiation-induced damage.
The results of phase I are presented in Sec. 6.

\textbf{Phase II:} For the second phase of radiation testing, the
CCD201-20 will remain operational and will be maintained
at cold operating temperature ($-108^{\circ}$ C; $\sim$165K) while being
subjected to multiple DDDs with a cumulative dose equal to
the 6-year mission life at L2. The device performance will be
characterized before and after four separate proton doses in
order to provide information on the rate of performance degradation
throughout the 6-year lifetime. The results of phase II
testing will provide a worst-case test of radiation hardness, representing
flight operation over 6 years without the benefit of
in-flight warm cycling or any longer term annealing that may
occur over time at this colder temperature.  

We note that although Michaelis et al.\cite{michaelis13} carried out a DDD
study of the CCD201-20, the device was operated at
5 frames per second for signal levels of 10 e$^{-}$, with multiplication
gain of 100. Since the effects of radiation can greatly affect
a device in different ways based on its operating conditions
(discussed later in Sec. 6), we conducted the phase I study as
a survivability test for the CCD201-20 specifically for the
WFIRST-CGI. Therefore, the sensor was run under higher gain
conditions (by a factor of 2), for both high flux ($>$1600 e$^{-}$) and
low flux ($\sim$8 e$^{-}$) signals, and for an integration times of 100 s.
This test plan was important based on the disparate operating
modes of the CGI. In phase II, the CCD201-20 will be irradiated
over the full range of proton fluences reflecting beginning of life
(BOL) to end of life (EOL) for L2, while kept at cryogenic temperatures in the beamline, where the device will be under
power at all times in order to measure flat-band shifts as
expected in flight. This will be the first time that a CCD201-20 is subjected to this kind of radiation test.

\subsubsection{Performance optimization}

Proton radiation results in the creation of charge traps, which
reduce the EMCCD charge transfer efficiency (CTE). The
CTE can also be written as 1$-$CTI, which is the ``charge transfer
inefficiency''. We will use the CTI term for discussion of this
phenomenon for the remainder of this paper. Techniques such as
pocket pumping\cite{janesick01,hall14a, hall14b, hall14c,wood14,murray12b, murray12a} will be used to characterize the traps in the
postirradiation EMCCDs. Once characterized, various operational
mitigation techniques, such as charge injection, and various
clock waveforms will be used to optimize the performance
in the presence of charge traps\cite{gow12}. In addition, postprocessing
correction algorithms\cite{massey14} will be tested.
  
\subsubsection{Thermal environment testing}

The CCD201-20 will undergo multiple thermal cycles spanning
the mission survival temperature range and the resulting performance
degradation will be characterized. A device will also
undergo a 24-h thermal soak at the warm and cold survival
temperature extrema.

\section{ELECTRON MULTIPLYING CCDs AND THE CCD201-20}
\label{sec:ccd201}  % \label{} allows reference to this section

The introduction of the EMCCD\cite{jerram01,hynecek01} enabled subelectron effective
read noise performance over a wide range of readout rates.
Its architecture is very similar to that of a conventional CCD.
The difference lies in the EMCCD's so-called high-gain or
EM register. This is an extended multiplication stage containing
a large-signal output that lies after the conventional serial register.
In this region, electrons can be amplified by a process known
as avalanche multiplication, in which high-voltage clocks produce
electric fields in the silicon that are high enough to amplify
charge via impact ionization. Crucially, this process is inherently
stochastic, where there is $\sim1 - 1.5\%$ probability of an extra
electron getting generated per electron per given multiplication
stage. Since a device can contain hundreds of multiplication
stages, the probability of amplification becomes significant,
and the subsequent EM gain, G, can amplify the signal by factors
of thousands:

\begin{figure}[!t]
   \centering
   \includegraphics[width=14cm]{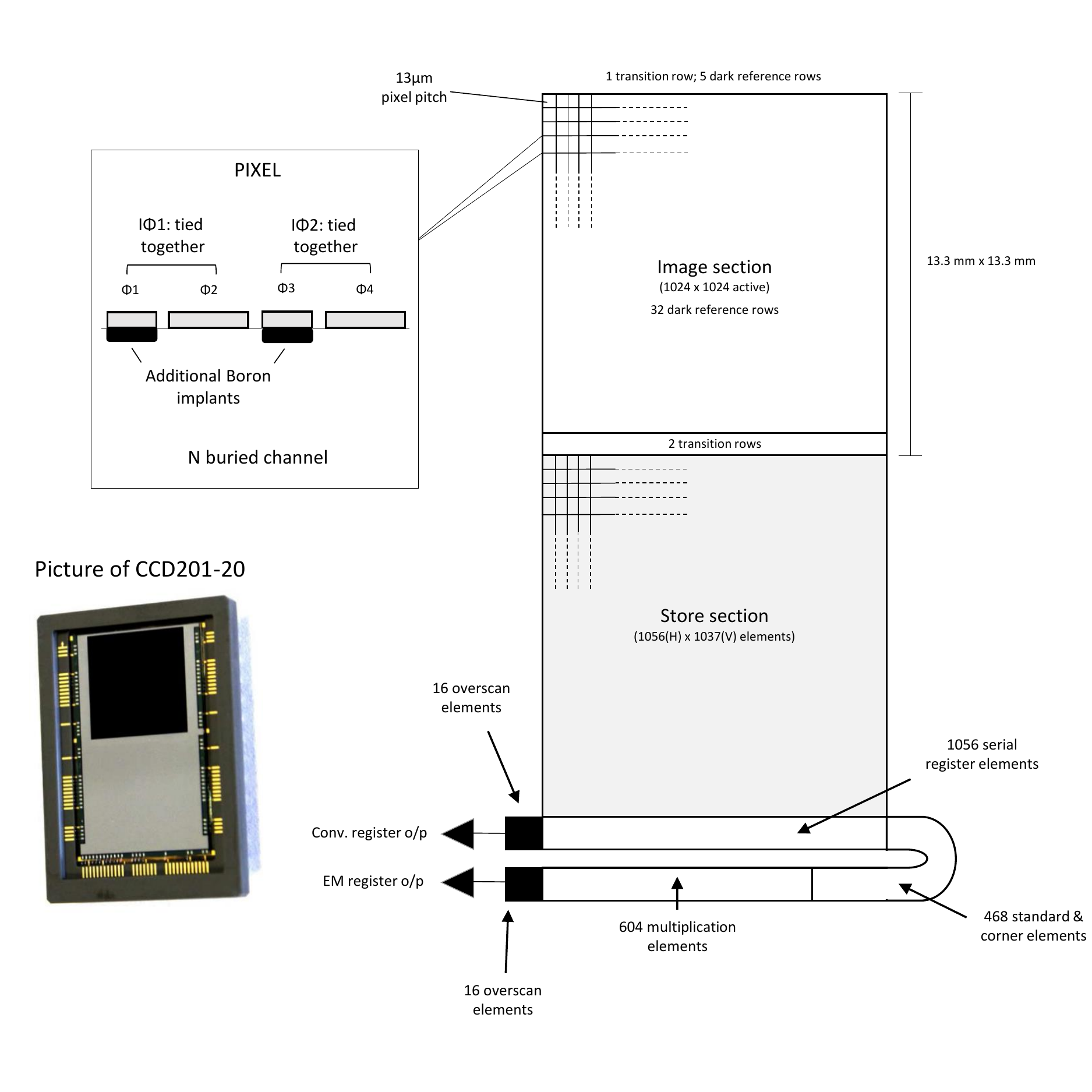}
\caption{Schematic of CCD201-20. Charge is passed from the image section to the store section and
subsequently read out through either the conventional or EM registers. To the left, we also include
the approximate configuration of the two-phase boron implants within a pixel, where the image and
store sections have an identical design. Electrode pairings, as indicated, is employed for standard
clocking. Adapted from the e2v CCD201-20 specifications sheet.}\label{fig:CCD201}
\end{figure}

\begin{equation}\label{eq:gain}
G=(1+\alpha)^N,
\end{equation}

where $\alpha$ is the $\sim$$1 - 1.5\%$ probability of amplification and $N$
is the number of stages. The result is a much higher S/N, albeit
at the proportional reduction in pixel charge handling capacity.
It is important not to continuously saturate the device under
high amplification, since this can reduce the lifetime of the
EM register. We note that there is an added variance in the
EM output that consequently reduces the S/N, which results in
an effective reduction of the sensor's QE by up to $\sim$50\%. This is
referred to as the ``excess noise factor'' (ENF); however, post-readout
techniques can reclaim this reduction in QE\cite{daigle08}. The
ENF can be calculated as follows:

\begin{equation}
ENF^{2}=\frac{\sigma_{op}^{2}}{(G)^{2} \cdot \sigma_{ip}^{2}},
\end{equation}
where $\sigma_{op}^{2}$ is the variance in the output and $\sigma_{in}^{2}$ is the variance in the input (Robbins \& Hadwin 2003\cite{robbins03}).  The ENF asymptotically approaches $\sqrt{2}$ for gains $\geq$10, as shown:

\begin{equation}\label{eqENF}
ENF^{2}=\frac{2}{\alpha + 1},
\end{equation}
because $\alpha<<G$.  This is the theoretical limit of the noise floor.  

\begin{table}
\caption{Specifications of the CCD201-20 EMCCD, from e2v.}
\begin{center}
\begin{tabular}{cc}
\hline
Parameter & Specification \\
\hline \hline
Sensor family & EMCCD \\
Variant & BI$^{\ast}$, 2-Phase \\
Active pixels (image) & 1024 (H) $\times$ 1024 (V)\\
Frame Transfer (store) & 1056 (H) $\times$ 1037 (V) \\
Image area & 13.3 mm $\times$ 13.3 mm \\
Pixel pitch & 13 $\mu$m \\
Active area CHP$^{\dag}$ & 80,000 e$^{-}$ pix$^{-1}$\\
Gain register CHP$^{\dag}$ & 730,000 e$^{-}$ pix$^{-1}$\\
Fill factor & 100\% \\
\# O/P amplifiers & 1 $\times$ Conv., 1 $\times$ EM \\
Multiplication elements & 604 \\
Dark reference columns & 32 \\
Overscan elements & 16 \\
\hline\\
\multicolumn{2}{c}{$^{\ast}$BI = Back-Illuminated; $^{\dag}$CHP = charge handling capacity.}\\
\hline
\end{tabular}
\end{center}
\label{table:CCD201}
\end{table}

The high gain amplifier is driven by much higher voltages
($>$40 V) than the conventional register ($\sim$11 V) and can deliver
an effective read noise of $<<$1 e$^{-}$ rms with EM gain, whereas the
conventional amplifier provides a typical noise of $\sim3 - 10$ e$^{-}$ rms. Importantly, as a result of much higher bias voltages
applied via the high-voltage clock (that directly controls the
gain), relatively low amounts of raw signal can quickly lead
to saturation of pixels in the high-gain register. Therefore,
there is always a tradeoff between the desired reduction in
read noise and a reduction in the effective pixel charge capacity
in order to achieve the highest S/N yield. Indeed, the employment
of high gain at low temperatures (less than approximately $<$ -90$^{\circ}$ C) can also affect a sensor's CTI, which is discussed later
in Sec. 6. Thermal stability in the CCD controller is extremely
important, since the gain will remain fixed for a fixed CCD temperature,
T but will strongly vary as a function of $\delta T$ but also
for changing bias voltage conditions. All of these factors, in
addition to those discussed later in this document, must be
heavily considered when optimizing an EMCCD for low-flux
observing. By considering the high dynamic range provided by
the conventional amplifier and the high sensitivity provided by
the EM amplifier, EMCCDs are placed as an ideal detector for
many astrophysical applications, including the WFIRST-CGI.

As previously outlined in Sec. 1.2, the CCD201-20 sensor
(hereafter CCD201) has been baselined for the CGI, for
both the Imager and the IFS. Although development of a
$\sim$4K $\times$ 4K EMCCD is under way (the e2v CCD282), it is
still at the early stages of testing\cite{gach14}, thus making the CCD201
the largest-format EMCCD device currently available from
e2v. It is a two-phase, frame transfer sensor, and has dual output
amplifiers, as outlined in Table 5 and shown in Fig. 3. Even
though the CCD201 is designed to be operated as a two phase
CCD, four-phase clocking is possible because each of
the parallel clocking lines are connected to a separate pin on
the package; however, e2v has included two boron implants
under phases $\phi$1 and $\phi$3, thus facilitating IMO, which, in addition
to deep sensor cooling, further suppresses contribution from
minority carriers from thermal generation, namely surface dark
current. We have used this capability for four-phase clocking of
the CCD201 in order to use the trap pumping technique\cite{janesick01} to
characterize radiation-induced traps (see Sec. 6.3). All of the
results that are outlined in this paper were carried out with standard
back-illuminated silicon CCD201s, which have an image
section of 1024 $\times$ 1024 active pixels with a store section of
1056 $\times$ 1037 pixels, see Fig. 3. Thinning and back-illumination
help to improve QE, which peaks at $>$90\% at 550 nm (we currently
use the standard midband e2v coating\cite{e2v}. The CCD201
can demonstrate a maximum well depth of 80,000 e$^{-}$ in the
image section and 730,000 e$^{-}$ in the gain register. Pixels in
the gain register are designed to have a large full well capacity
(FWC) in order to avoid saturation during readout. The full
spectral range ($>$15\% QE transmission) of the sensor lies
between $\sim300 - 1000$ nm, allowing tests in all of the
WFIRST-CGI Imager and IFS bands.

\begin{figure}[!t]
   \centering
   \includegraphics[width=12cm]{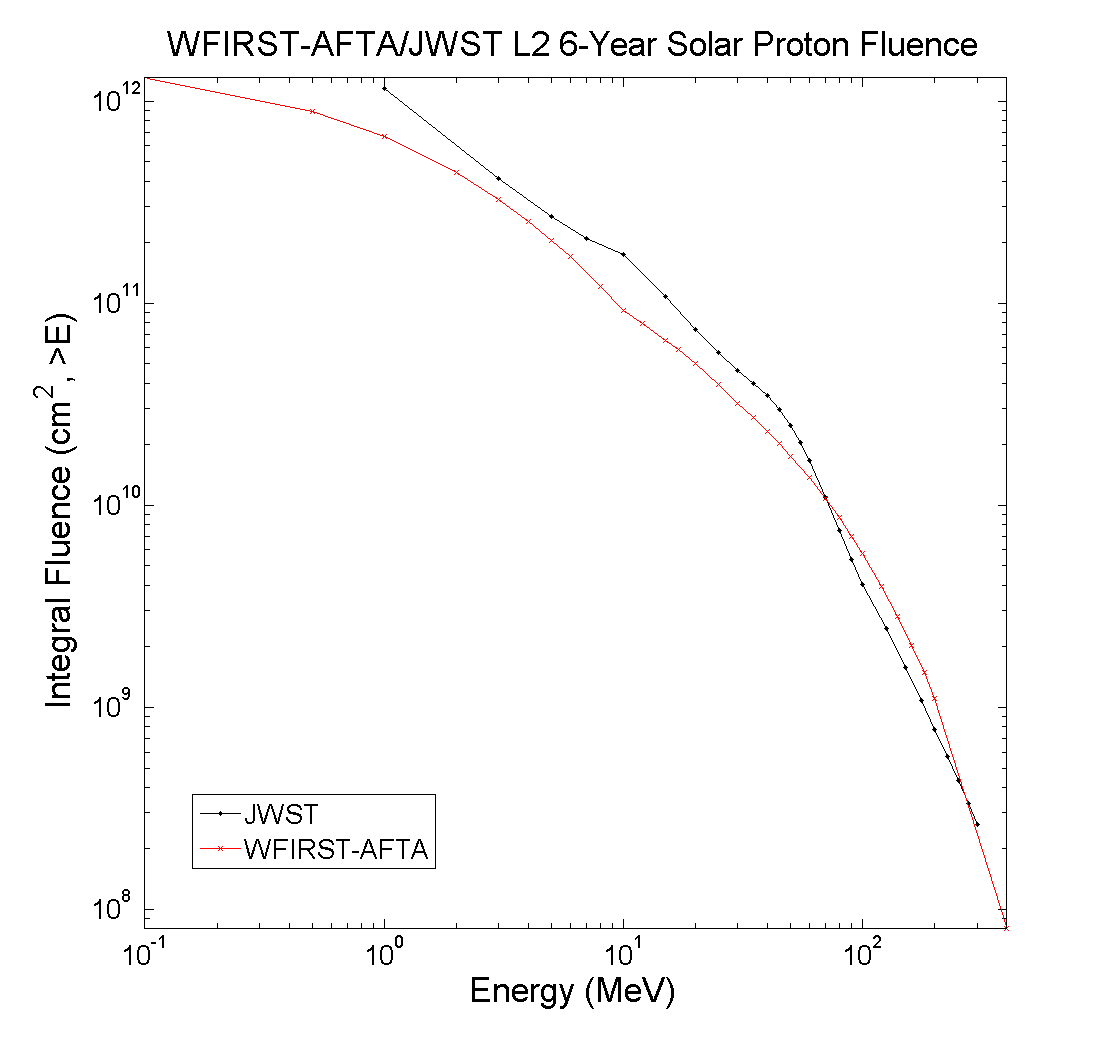}
\caption{Comparison of independent predictions for the solar proton
fluence in a direct insertion L2 orbit for the WFIRST and JWST
missions. WFIRST data were calculated based on the JPL 91 Solar
Proton model at a 95\% confidence level and with a radiation design
factor (RDF) = 2. JWST data were scaled to 6 years based on 5-year
data taken from ``The Radiation Environment for the JWST'' (JWSTRPT-
000453)\cite{barth03}.}\label{fig:rad-L2-geo}
\end{figure}

\begin{figure}[!t]
   \centering
   \includegraphics[width=12cm]{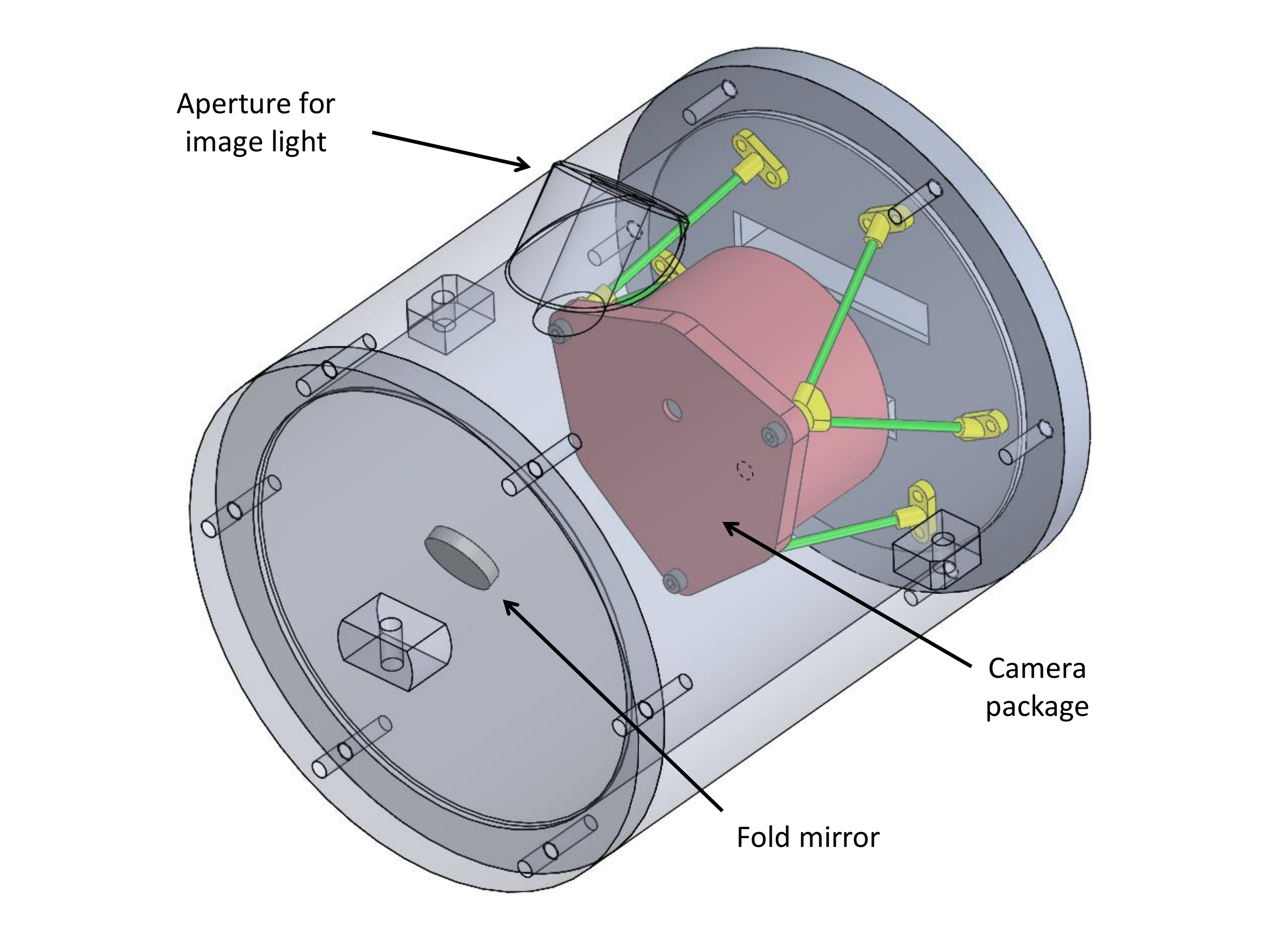}
\caption{Computer-aided design (CAD) model for the detector package
and radiation shielding, shown 10-mm thick. The shielding was
designed for radiation analysis. The camera package is mounted
inside a cylindrical drum radiation shield as shown. The radiation
simulation compared different thicknesses of both tantalum and
aluminum shielding material.}\label{fig:shield}
\end{figure}

\section{WFIRST SPACE ENVIRONMENT}
\label{sec:wfirst}  % \label{} allows reference to this section

As part of the current design cycle (cycle 6), the WFIRST
project is carrying out a trade between a geosynchronous
orbit and an L2 orbit similar to that of the James Webb
Space Telescope (JWST). If the project selects the L2 orbit,
then it is most likely to be a direct insertion orbit whose trajectory
through the Earth's trapped-particle radiation belts will
be very short and inconsequential. Therefore, the exposure to
harmful radiation in general is lower at L2 compared to geosynchronous
orbit.

To calculate the radiation environment for the WFIRST-CGI
detector in L2, we ran the radiation transport computer code for
a geosynchronous orbit and subsequently removed the contribution
from the Earth-trapped protons and electrons, leaving only
the solar proton fluence, common to both geosynchronous and
L2 orbits. To verify this approach, a comparison was made
between the JPL model prediction of the L2 environment for
WFIRST and the Goddard Space Flight Center (GSFC)
model prediction of the L2 environment for the JWST mission.
Since different NASA centers use different validated radiation
models, the comparison serves as an independent corroboration.
The graph in Fig. 4 shows the solar proton fluence as a function
of particle energy for WFIRST and JWST. The WFIRST model
prediction is only slightly lower than that of JWST. The
WFIRST data were calculated based on the JPL 91 Solar Proton
model at a 95\% confidence level and with a radiation design
factor (RDF) of 2. JWST data were scaled to 6 years based
on 5-year data taken from ``The Radiation Environment for
the JWST'' (JWST-RPT-000453)\cite{barth03}. Based on this comparison,
it was concluded that our methodology was sound.

A radiation transport model, NOVICE (Jordan et al. 2006\cite{jordan06}), was used to calculate DDD and total ionizing dose (TID) in the IFS EMCCD die. A three-dimensional (3-D) mass model of the EMCCD package
and radiation shield computer-aided design (CAD) (shown in
Fig. 5) was used for the radiation transport calculation.
Shielding is essential to limiting radiation-induced damage.
To assess the DDD and TID in the CCD die, seven ``dose detectors''
along the CCD die center perpendicular line were selected.
Thus, the dose levels were calculated at the detector surface and
bulk depths of 2.54, 10.16, 25.4, 101.6, 254, 508, and 640 $\mu$m.

The radiation transport model revealed that there is a significant
difference between geosynchronous and L2 orbits for the
damage due to TID, whereas there is negligible difference
between the two orbits for the damage due to DDD. Using the
radiation model, it was determined that a 1-mm-thick glass window
in front of the EMCCD reduced the TID level in the surface
layer to only 1 krad (RDF = 2) for a radiation shield of 10-mm thick
tantalum. The radiation transport calculation incorporated
a 3-D CAD mechanical model of the camera package enclosed
by a cylindrical radiation shield, as shown in Fig. 5.

Figure 6 shows the expected radiation exposure of the
WFIRST-CGI detectors over the span of a 6-year mission in
geosynchronous and L2 orbits, for both DDD and TID, respectively,
for a range of tantalum and aluminum shielding thicknesses
and including a 1-mm-thick glass window placed in front of the detector. The model used the required RDF of 2.
The design for the EMCCD shielding material and thickness
will later be refined based on the results of the phase II radiation
testing to be conducted in the summer of 2015. The DDD radiation
doses to be used in phase II will span the radiation levels
across the family of curves shown in Fig. 6. In phase II, we will
measure the detector performance degradation as a function of
DDD fluence. Then, the shielding material and thickness will
be chosen to correspond to an acceptable EOL performance
degradation.

Based on the above radiation analysis, it was concluded that
so long as a 1-mm-thick glass window is used in front of the
detector, TID will be reduced to very low levels and DDD is
the primary risk to EOL detector performance.

\begin{figure}
   \centering
   \includegraphics[width=17cm]{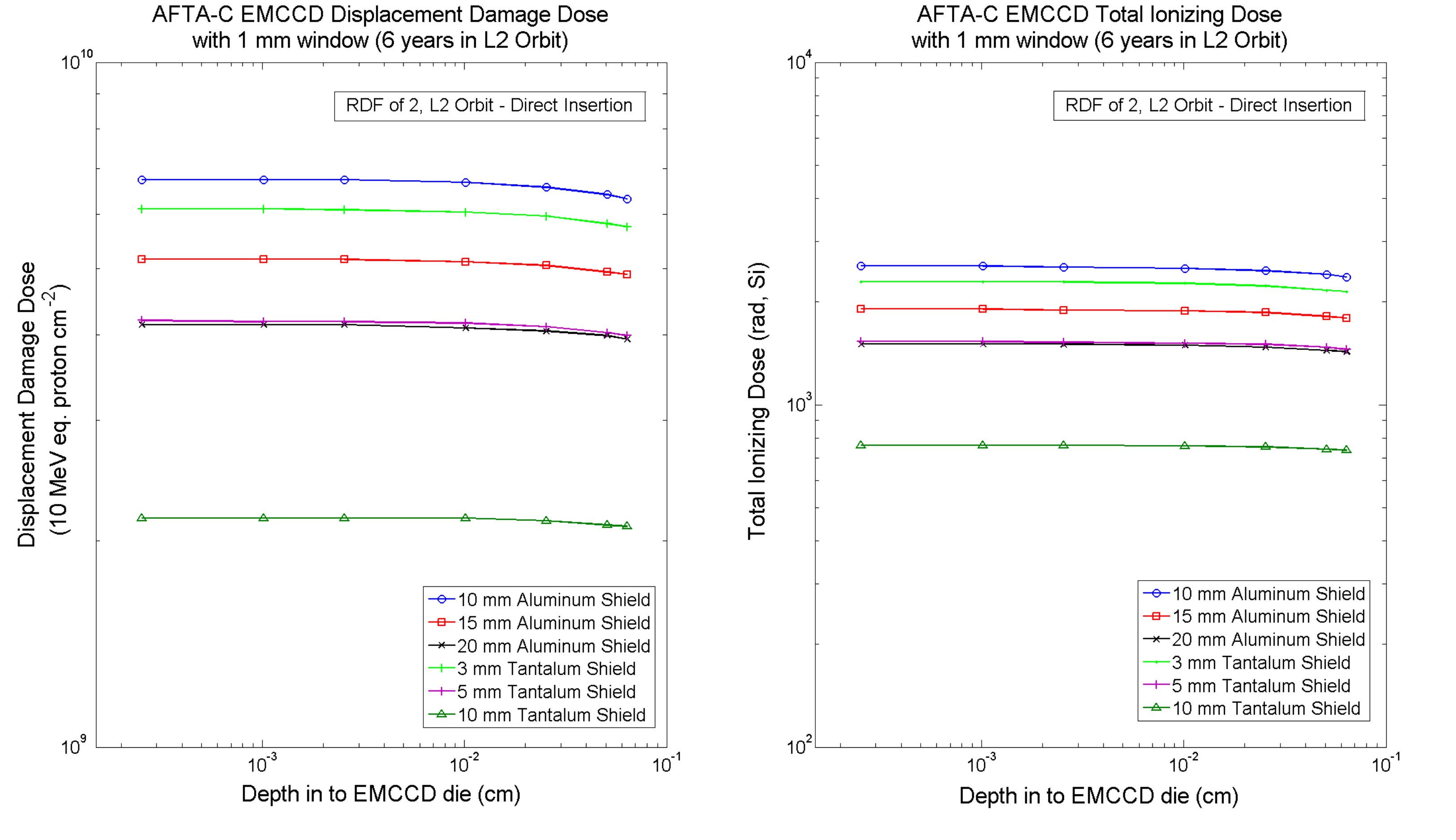}
\caption{\textit{(Left)} Logarithmic plots of DDD (10 MeV equivalent protons cm$^{-2}$) as a function of depth in the
EMCCD (cm) with a 1-mm-thick glass window. This is for an L2 orbit (direct insertion) for an
RDF = 2. DDD is shown for a range of various shielding materials and thicknesses. \textit{(Right)} Logarithmic
plots of TID (rad, Si) as a function of depth in the EMCCD (cm) with a 1-mm-thick glass window.
This is for an L2 orbit (direct insertion) for an RDF = 2. TID is shown for a range of various shielding
materials and thicknesses.}\label{fig:rad}
\end{figure}

\section{RADIATION DAMAGE}
\label{sec:radiation}  % \label{} allows reference to this section

\subsection{EMCCDs for the WFIRST-CGI}

CCDs have been extensively developed and applied for precision
measurements in space, beginning with NASA's Galileo
spacecraft and the HST, and continuing to the present with large
focal planes on Kepler, Gaia, and Euclid. Radiation environments
in space and their effects on detectors have been extensively studied, and NASA/JPL have developed tools for modeling and
testing the effects of such environment on CCDs\cite{janesick01}. Despite the extensive heritage of silicon
detectors in space, there is at present no flight-proven detector
technology that meets WFIRST-CGI requirements.

As outlined in Sec. 1.2, the JPL detector trade study relative
to the WFIRST-CGI stringent requirements for sensitivity and
S/N identified that the CCD201 proved to be the most promising
detector in meeting such performance. However, for the
purpose of comparison, this study assumed BOL performance
only where degraded sensor performance, a consequence of
radiation damage in the form of TID and DDD, was unknown,
since EMCCDs had not yet been empirically tested for such
degradation.

In this context and in order to advance the TRL of EMCCDs,
it is therefore critical to understand the risks of radiation damage
and their mitigations. Although there are some previous studies
of radiation effects on CCD201 detectors,\cite{michaelis13} these studies are
not sufficient to demonstrate the EOL performance in the
WFIRST environment or its projected measurement conditions.
Consequently, the WFIRST-CGI requires experimental verification
of EOL performance of CCD201 detectors in a realistic
radiation environment.

\subsection{Radiation Damage Effects on Detectors}

In normal operation, light entering a CCD will interact with the
silicon to produce electron-hole pairs. CCDs form images by
first collecting photogenerated charge over a period of time
(charge storage) and then measuring the accumulated charge
as a function of location on the detector (readout). Latent images
comprise accumulated charge stored in capacitors within the
array, and images are ``read'' by serially transferring collected
charge through the coupled capacitors in the array toward an
output amplifier that converts charge packets to an analog voltage
output signal. EMCCDs further incorporate a high-gain
register, which uses high electric fields to amplify the charge.
Maintaining CCD performance within specifications depends
on charge trapping and detrapping properties of silicon, oxides,
and silicon-oxide interfaces, all of which are subject to degradation
due to radiation exposure.

Silicon detectors exposed to high-energy photons, electrons,
and particles suffer damage to both the silicon and the oxides
that together comprise the capacitors making up the CCD structure.
Particles and photons incident on the array transfer energy
to the CCD along their paths, leaving behind damage in the form
of broken chemical bonds and charged defects. Occasional interactions
with nuclei can further damage the detector by displacing
silicon atoms from their ordered locations within the crystal
lattice. These two types of damage--ionization and displacement
damage--create defects and traps that affect the detector
performance in distinct ways.

\begin{figure}
   \centering
   \includegraphics[width=13cm]{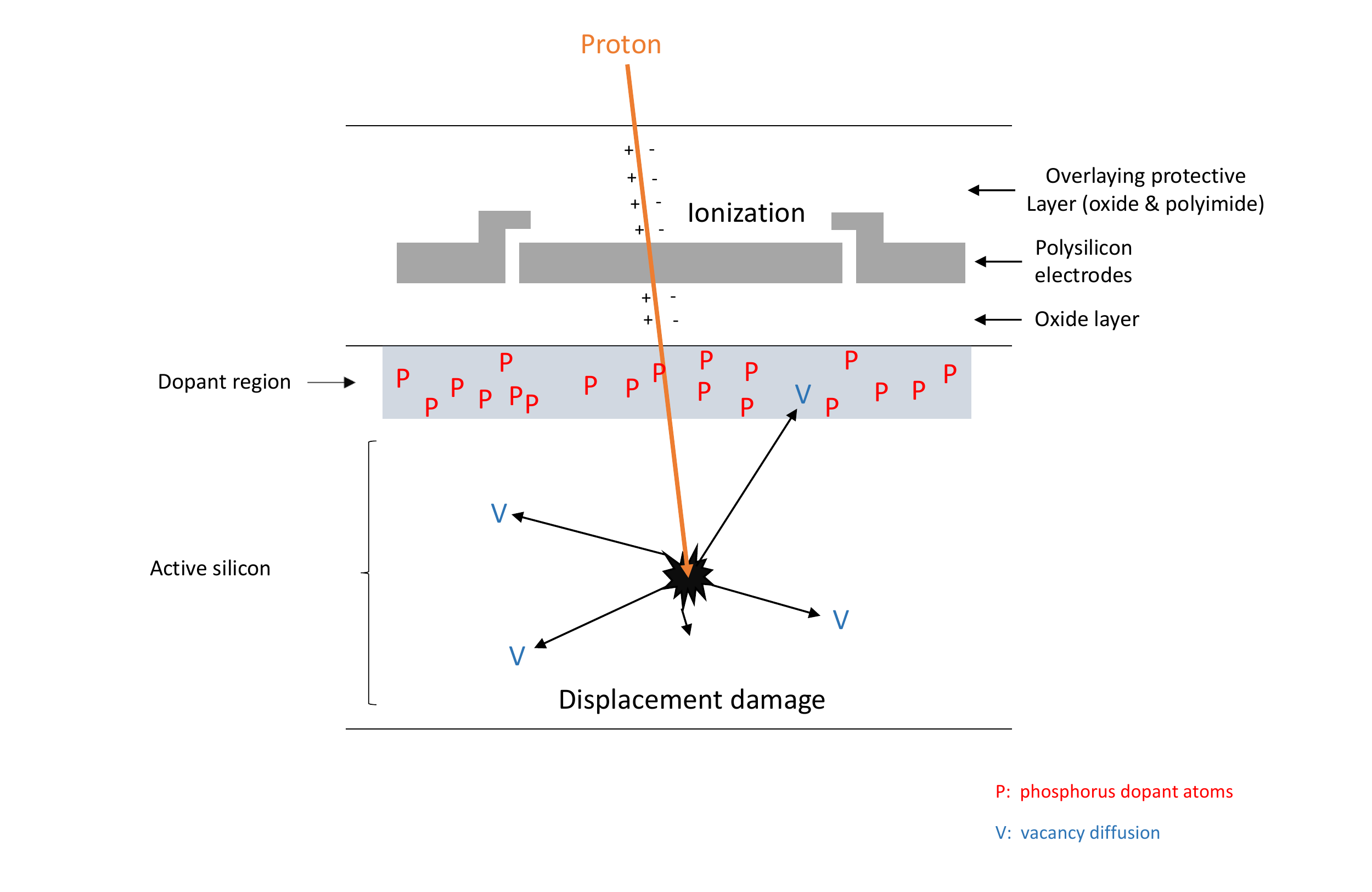}
\caption{Illustration of radiation damage mechanism for a high energy proton in an EMCCD, much like the damage that the CCD201s have undergone in this study.  Adapted from Hall et al.\cite{hall14d}.}\label{fig:DDD}
\end{figure}

\subsubsection{Total ionizing dose}\label{sec:TID}

Charged particles, x- and gamma radiation in the space environment
damage oxides and surfaces, leaving a trail of broken
chemical bonds in silicon oxides and interfaces. The resulting
defects act as electron and hole traps, leading to time-variable
charging of the oxide and the silicon-oxide interface, which, in
turn, affects the internal electric fields in the silicon detector. The
resulting ``flat-band voltage shifts'' alter the operation of CCDs
by changing the internal electric fields that direct and guide signal
electrons in their path to the output amplifier. Shifting bias
voltages can degrade detector performance by altering FWC,
CTI, and CIC. In an EMCCD, radiation-induced flat-band voltage
shifts can degrade multiplication gain. Radiation-induced
traps formed near the thinned, illuminated surface of the CCD201 are subject to variable (time-dependent) charging,
resulting in decreased QE, instabilities and image artifacts,
including persistence and QE hysteresis. Interface traps are also
responsible for increased surface dark current and loss of signal
due to surface recombination. Moreover, the oxide and interface
traps that are responsible for these effects can change with time,
which introduces time and temperature dependencies and performance
instabilities in radiation-damaged devices.

The severity of the radiation-induced degradation depends on
the oxide thickness and composition, as well as the processing
parameters used to form the oxide. e2v has developed radiation-hardened
oxides to mitigate these effects, reducing the flat-band
voltage shift from the typical values of 100 to 200 mV/krad (Si)
in standard devices to 6 mV/krad (Si) in devices fabricated
using radiation-hardened oxides.\cite{burt09} However, EMCCDs have
not been fabricated using a radiation-hardened oxide and there
would be some technology development and associated risk
required to do so. A radiation-hardened oxide is complicated
by the unusual conditions required to operate the multiplication
register. We suggest that the most effective design modification
for improvement of radiation hardness would be a reduction in
the charge channel width, which would reduce the cross section
for radiation-induced traps.

\subsubsection{Displacement damage}

\textit{Displacement damage and trapping.} Energetic particles
such as protons and neutrons can damage detectors by displacing
silicon atoms from their lattice sites in the silicon crystal (see
Fig. 7). Silicon vacancies are associated with energy levels in the
silicon bandgap, which can capture photogenerated electrons
and holes. One effect of traps is therefore a degradation of detector
sensitivity and increased noise, as trap-assisted generation and recombination events cause a loss of signal and increase
the bulk dark current. The greatest impact of displacement
damage on detector performance comes from the smearing of
charge that results from trapping and subsequent detrapping of
electrons, leading to artifacts, such as image persistence and
deferred charge.\\

\textit{Trap species and characterization.} Isolated vacancies in
silicon are unstable, and vacancies will migrate through the crystal
until they interact with other defects to form a stable configuration.
The most common trap species in radiation-damaged
silicon are divacancies, phosphorus-vacancy complex, and oxygen-
vacancy complex. Each of these forms a distinct set of trap
states in the silicon. Table 6 shows these common radiation-induced
defects and traps with their associated energy levels.

Once the trap properties are known, the dynamics of electron
trapping and detrapping events can be modeled using Shockley-Read-Hall theory.\cite{shockley52} The degradation of CCD performance
depends not only on the density, location, and type of traps
present in the detector but also on CCD operating parameters,
such as detector temperature, bias voltages, and clock frequencies.
If these parameters can be measured and/or calibrated,
it becomes possible to correct for trap-induced errors using
image postprocessing algorithms. Deferred charge can be corrected
on the HST to a precision of 97\%, and work is ongoing
to improve correction algorithms to achieve 99\% accuracy for
the ESA Euclid mission. However, image correction algorithms
run into limitations at low signal levels. For example, even with
perfect knowledge of the trap locations and properties, capture
and emission times for trapping/detrapping events are stochastic,
so that the accuracy that can be achieved by image correction
algorithms depends on the number of photons detected\cite{hall14a, hall14c,massey14,israel14}.

\begin{table}
\caption{Most common types of traps and physical trap parameters in
silicon.\cite{hall14d} These reflect a sensor temperature of --114$^{\circ}$ C. We refer the
reader to Hall et al.\cite{hall14d} for a discussion on trap cross section and emission
time. These must be considered when minimizing CTI based on
the dwell time of a charge packet under an electrode.}
\begin{center}
\begin{tabular}{ccc}
\hline
Trap & Defect & Average energy relative \\
 & & to conduction band \\
 & &  (eV) \\
\hline \hline
Si-A & Oxygen-vacancy complex & 0.17 \\
(V-V)- - & Divacancy, doubly ionized & 0.21 \\
Unknown & Unknown &0.30 \\
(V-V)- & Divacancy, singly ionized & 0.41 \\
Si-E & Phosphorus-vacancy complex & 0.46 \\

\hline
\end{tabular}
\end{center}
\label{table:traps}
\end{table}

\subsection{Associated Risks}\label{risks}

\subsubsection{Radiation-induced degradation of detector noise}

The CCD201 noise floor must be sufficiently low over the lifetime
of the mission that the WFIRST-CGI can detect the faint
signals from planets orbiting distant stars. Because detector
noise degrades over time due to radiation exposure, meeting
this requirement will entail measuring several key detector
parameters as functions of radiation exposure. Among these
key parameters are dark current, CIC, multiplication gain, and
read noise. In order for these results to be applicable for the
CGI, CCD201 noise performance must be characterized using
operating parameters that are relevant to a realistic environment
for the mission, including any background signals that may
affect low-light level sensitivity. In particular, the CR background
is a threat to low-light level detection on the CGI because of the
potential for CRs to contribute to the noise floor. While it is possible
to subtract CRs from images during image postprocessing, residual noise from this subtraction (such as shot noise) could
dominate the detector noise in the affected pixels. As a result,
the CGI detector frame rate must be kept short enough to prevent
the CR background from dominating the noise floor. It is
expected that the CGI will be required to collect and coadd
multiple frames during the instrument integration time, using
postprocessing to remove CRs from individual image frames.
In addition, the large-signal events generated by CRs, coupled
with the finite well capacity of the high-gain register, are
expected to impose a constraint on the maximum usable gain
setting. Finally, the EMCCD gain degrades over time (through
a process commonly known as ``aging''), and this degradation
must be characterized under realistic conditions because the
EMCCD gain directly affects the EMCCD noise floor.

\subsubsection{Cosmic ray \& other particle rates (ground-based)}

We conducted a study at JPL to estimate the number of CR
events at ground level that an EMCCD under high-gain conditions
would be sensitive to. This study was carried out in order
to investigate the optimum WFIRST-CGI integration times
based on these particle densities with the view to ultimately
investigate their effect on the device CTI. We found that
when using conventional CR identification and removal algorithms,
the CR core (which we define as the area where
the main strike occurred) was easily identified; however, we
observed that in the event that such a particle had sufficient
energy to saturate the high-gain register (after EM amplification),
the algorithms were unable to fully identify pixel regions
that were affected by this charge. This charge appears as a
streak, or ``tail'', in the direction of charge transfer. In this
study, we sought to identify the particle rate and the fraction
of pixels affected by considering CR cores as well as CRs
and tails. The algorithms presented in this work were designed
to identify these regions only; they do not perform any postprocessing
techniques to mitigate these effects on CTI. This will
be considered at a later time.

\begin{figure}
   \centering
   %\hspace{2em}
   \includegraphics[width=15cm]{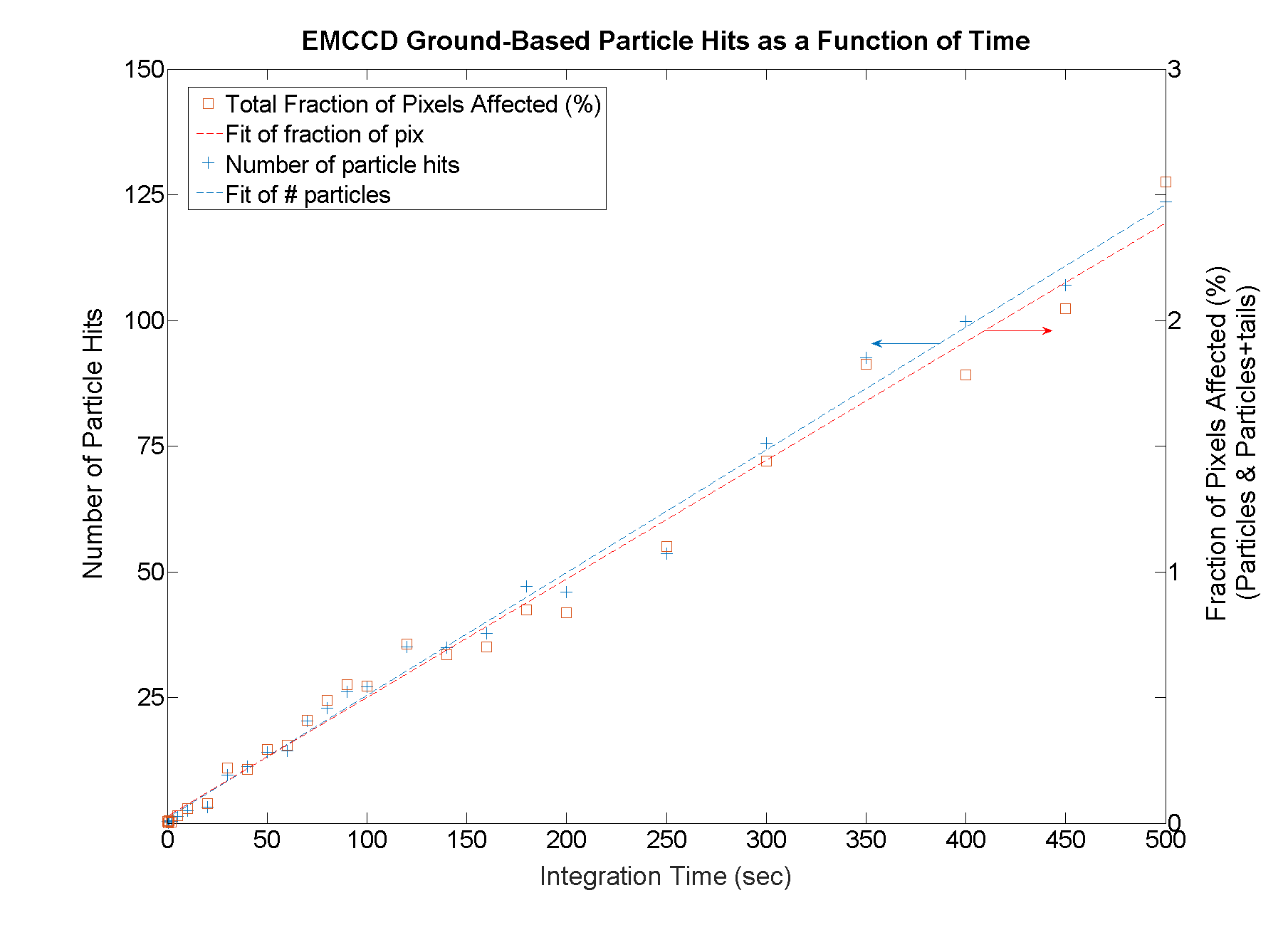}
\caption{Plot of the number of particle hits (y-axis, a) as a function of integration time (s), and the total
fraction of pixels affected (\%, y-axis, left) as a function of integration time (s). The horizontal arrows point to
which \textit{y-axis} is relevant for each set of data. For this test, we selected acquisition settings that meet
the WFIRST-CGI requirements, running at a temperature at −85$^{\circ}$ C and using an EM gain of 450.
The total fraction of pixels reflects the full image area of 1024 $\times$ 1024 pixels.}\label{fig:CRs}
\end{figure}

We calculate a rate of 0.148 particles cm$^{-2}$ sec$^{-1}$ for the CCD201 (13-$\mu$m pixel pitch) and show these data in Figure~\ref{fig:CRs}. We found this by selecting appropriate acquisition settings
that reflect the CGI conditions in order to meet dark current,
CIC, and effective read noise requirements (EM gain of 450
was used), and exposed the sensor for integration times of
0 to 500 s, yielding as many as $\sim$125 events for a 500-s exposure
where the sensor was set up such that its orientation was parallel
to the ground with no shielding. Each data point was calculated
from 10 frames where the average number of events was calculated
in addition to the fraction of pixels that were affected by both an event and/or an event and a streaked signal, or tail.
Here, we refer to an event as a CR, or other particles that the
EMCCD was sensitive to at these gain levels.

In order to identify and/or remove particles in the EMCCD
that could increase device CTI, we developed an in-house detection
and removal algorithm using the Python language from
the Python Software Foundation.\cite{python} We show a flowchart of
this procedure in Fig. 9. The identification algorithm has
three different classes: ``Image Handler'', ``Section Group'', and
``Section''. The ``Image Handler'' class first calculates the
average pixel value in an image, identifies pixels above some user-defined threshold (this variable changes the aggressiveness
of the algorithm), and assigns a flag to these pixels. Once these
pixels have been flagged, they are broken up and distributed into
“Sections” by the “Section Group” class. Here, we define a section
as an array of pixel coordinates corresponding to flagged
pixels as well as all adjacent pixels, e.g., those identified by
the “Image Handler.” Each section is then sorted based upon
two variables: (1) the size of a section in units of pixels and
(2) the elongation of an event (e.g., a nonsymmetric core or
a core and its tail). Each variable can change the aggressiveness
of the algorithm and is user defined. The first variable checks
how many pixels are in a section and the second variable defines
the minimum required dimension ratio (length:width) in order
for a section to be flagged as an event and a tail. Therefore,
a section needs to be at least N times larger in one dimension
than all of the other dimensions. These dimensions reflect a
square or rectangular box, with eight reference points (N, S,
E, W, NE, NW, SE, and SW). By establishing these points,
the algorithm identifies the shape of a section. We found that
a minimum length to width ratio of 1.5 was sufficient to identify
nonsymmetric cores or cores and tails. The algorithm then takes
sections that have passed both of the requirements and flags
them as an event or events with tails. Finally, all flagged pixels
are replaced with a value of --300 in order to easily neglect during
image processing. This is also useful for estimating dark
current rates for longer exposure times since false-positives
are rejected. Finally, a pixel map is created where real pixels
have a value of zero and events have a value of --300.

\begin{figure}
   \centering
   %\hspace{2em}
   \includegraphics[width=17cm]{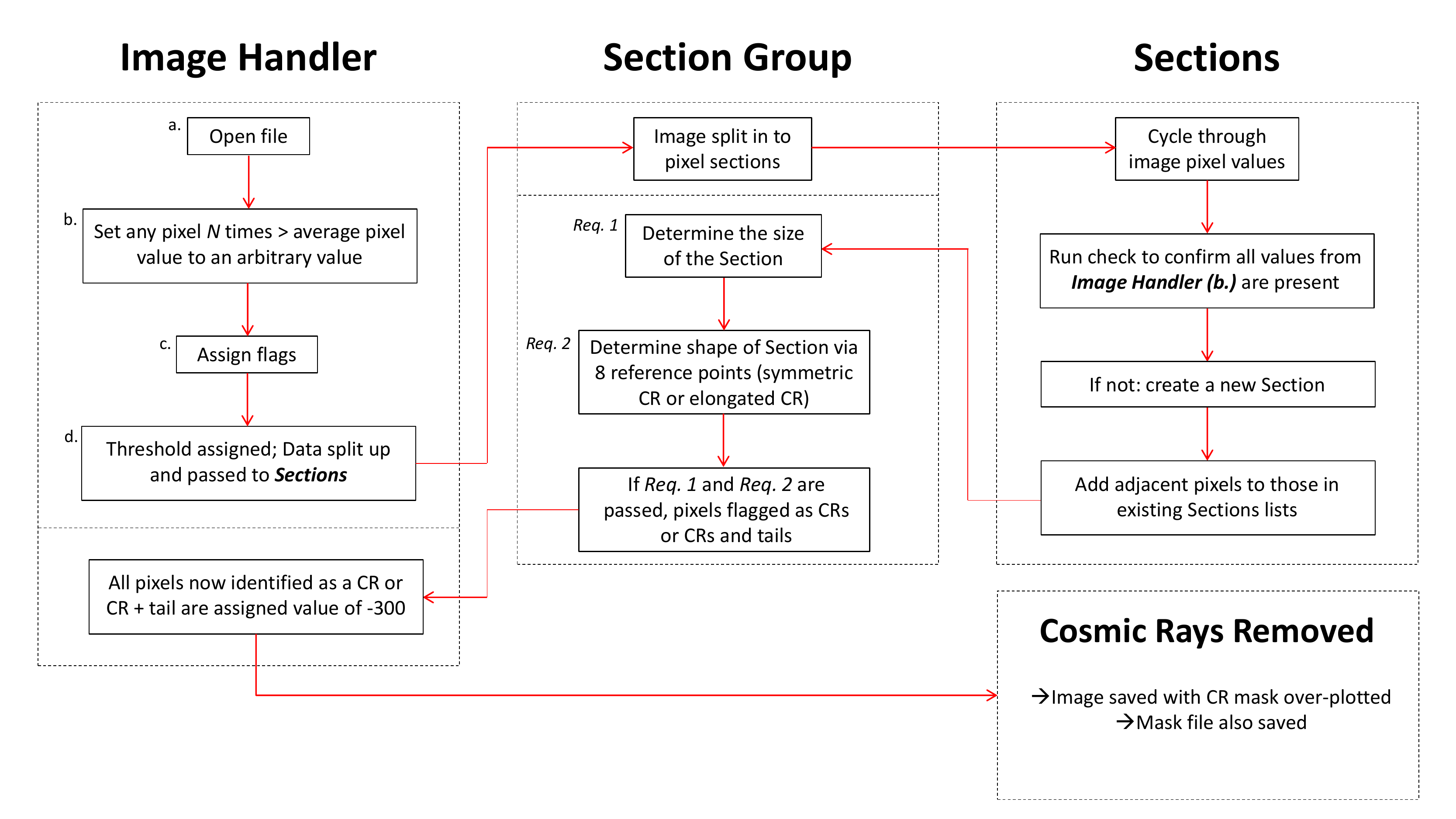}
\caption{Flowchart of the custom EMCCD cosmic ray/other removal algorithm that was developed at JPL for this work.  The three classes, \textit{Image Handler}, \textit{Section Group} and \textit{Sections}, define the structure of the algorithm.}\label{fig:CR-flowchart}
\end{figure}

We note that each event profile on the detector has unique
properties based on, e.g., the angle of the particle strike with
respect to the detector surface, the length of the ionization
trail, the device epitaxial thickness, the energy of the particle,
the integration time, the employed EM gain and thus the degree
of saturation in the high-gain register, etc. The depth of solar
minimum and other synodic factors must also be considered
for particle rates. Therefore, large frame-to-frame variation can
be observed. In one frame, for example, there may be a large
number of events that are tightly confined to smaller pixel
areas where saturation in the high-gain register only occurs
for a few pixels (number of events--higher; fraction of pixels
affected--lower). A different frame may contain less particle
hits per unit area but with greater energy that are spread over
many pixels as a result of saturation in the high-gain register
(number of events--lower; fraction of pixels affected--higher).
We note that this model does not take multiple particle strikes on
one single pixel into account.

Other missions with L2 insertions, such as the Herschel
Space Observatory, found record rates of CRs and other particles
at launch, which they attributed to the depth of solar minimum
at the beginning of the mission. These data were taken
with Herschel's Standard Radiation Environment Monitor
(SREM) where typical SREM count rates in L2 were $\sim$3 -- 5 particles cm$^{-2}$ sec$^{-1}$; this range represents particles over a
range of energies from $E>0.5$ MeV -- $E>39$ MeV. We
refer the reader to the SREM database\cite{SREM} for data on the CR
flux over the full mission, showing a higher concentration of
CR flux at the beginning due to the solar minimum as indicated
previously. These particle rates, coupled with the predicted
solar cycle for the WFIRST launch, in addition to the range
of particle energy levels in L2, must be considered for the
CGI detectors. The ground-based study presented in this section
indicates a small fraction of the expected particle rate in space. Our measured CR rate is higher than what is typically measured
on the ground ($\sim$0.025 particles cm$^{-2}$ sec$^{-1}$)\cite{janesick01}; however, ground-based
particle fluence can vary greatly with altitude, in addition
to other factors such as the presence of high-potassium BK7
glass causing secondary effects, whether a sensor is back-thinned
as well as how thick it is or sensitivity to particles of
different energies than other studies by using high EM gain.
Device orientation can also affect the rate. We do not calculate
energy levels or particle type in this paper but sought to identify
any event that could potentially increase CTI. CRs are therefore
included in this sample. In this respect, our study is being used
to estimate optimum single-frame integration times to minimize
CTI, currently estimated to be $\sim$100 -- 300 s. These frame
times may be reduced further based on higher particle rates
in orbit and the associated behavior of CTI. We intend to further
extend this work to investigate such particle populations and
how the acquisition time should be tailored to minimize CTI.
Development of postprocessing techniques for the WFIRST-CGI
detector application, such as those carried out on HST's
WFC3 detectors must also be considered.\cite{massey14}

\subsubsection{Radiation-induced degradation of detector signal}

The CCD201 must be capable of detecting faint objects even
when the flux is low enough that it is necessary to detect single-
electron signals in any given frame. Here, it is understood
that it will be necessary to coadd many individual frames in
order to achieve the required S/N. The major challenge is
the effect of radiation-induced traps on CCD readout. While
trap densities in ``as-manufactured'' detectors are exceptionally
low, radiation exposure will cause significant accumulation of
traps during the lifetime of the WFIRST mission. As described
in Sec. 3, a radiation shield can extend the lifetime of the
CCD201, but no amount of shielding can eliminate this accumulation
of traps over time. Single-electron signals are vulnerable
to a loss of spatial information as electrons can be trapped
during readout and later released. In the limit of detecting single electron
signals, it is an extreme challenge to obtain enough
information to accurately correct images for trapping/detrapping
events.

Recently, however, we point out that ESA, together with the
Center for Electronic Imaging (CEI), have carried out a Gaia
Radial Velocity Spectrograph (RVS) study, where test spectra
were propagated through the RVS CCD sensors, including
irradiated sections of the chips. They confirmed that a signal of
1.5 e$^{-}$ pix$^{-1}$ could survive, and subsequently verified that
subelectron detections were possible. This work is therefore
extremely applicable to the WFIRST-CGI, and will be considered
in the upcoming work with irradiated EMCCDs. Details of
projected radiation damage to these sensors can be found in
Seabroke et al.,\cite{seabroke08} and the study confirming subelectron detections
for CCDs that have been subjected to a high radiation
dose can be found in Dryer and Hall.\cite{dryer14}

\subsection{Single Event Gate Rupture in EMCCDs}

A single event gate rupture, or SEGR, is when a heavy ion
strikes a gate electrode of a MOS-type semiconductor device,
such as an EMCCD, which subsequently damages the region
where the gate lies. For a SEGR event, the magnitude of this
damage can depend on factors such as the ionization energy
of the particle or the design of the device. Crucially, the voltages
that operate in this local region where the incident ion strikes are also a major factor. Since an EMCCD's high-gain register can
produce much larger local electric fields (3 MV cm$^{-2}$ interpoly field) based on voltages approximately four to five times higher
than a conventional register, this effect is of special relevance to
an EMCCD device. See Sexton et al.\cite{sexton97} and references therein,
for more discussion on the SEGR mechanism.

A previous study of the effects of SEGR has been carried out
with an EMCCD sensor,\cite{evagora12} who used the CYClotron of LOuvain
la NEuve (CYCLONE) at the Heavy Ion Facility (HIF), to
investigate the resilience of EMCCD technology to the effects
described above. Heavy ions in the space environment pose a
risk of catastrophic failure in EMCCDs, because of the potential
for an SEGR event to create a conductive path through one of
the gate oxides in the gain register. In order to test this experimentally,
an e2v CCD97 was used, which is a smaller format
device compared to the CCD201, with an active image area of
512 $\times$ 512 pix and a 16-$\mu$m pitch. Simulations carried out
prior to this test indicated potential gate rupture at $20 - 40$ MeV cm$^{-2}$ mg$^{-1}$ and therefore linear energy transfers
(LETs) of 58.8 MeV cm$^{-2}$ mg$^{-1}$ were used (the highest LET
available at the HIF), with 16-$\mu$m penetrating depth. Different
flux levels and amplification gain settings were used; however,
the CCD97 did not undergo a catastrophic failure due to gate
rupture. The multiplication gain was then increased and the
angle of the HIF changed to produce an apparent LET of
135 MeV cm$^{-2}$ mg$^{-1}$ in an attempt to induce a gate rupture.
Again, the device survived the test.

This test was successful in demonstrating that there was
no sign of a SEGR event for an EMCCD under high-gain
conditions. Furthermore, a spacecraft in L2 orbit is subjected
to $\sim$10$^{6}$ heavy ions m$^{-2}$ over 5 years with an LET of
60 MeV cm$^{-2}$ mg$^{-1}$, and the fluxes tested in the Evagora
et al.\cite{evagora12} study were sufficiently high to conclude that the probability
of an ion incident on a sensitive region of the device
in orbit would be small. We note that SEGR events are still
possible in EMCCDs; however, this result suggests that even
by considering the higher local electric field conditions present
in an EMCCD under high gain, such a device should not
be affected by heavy ions in orbit over the WFIRST projected
lifetime. Finally, another study by Smith et al.,\cite{smith06} who subjected
100 CCD97 EMCCDs to a proton fluence of 2.5 $\times$ 10$^{9}$ protons cm$^{-2}$ (10 MeV equivalent), also reported no device
failures, where they were specifically concerned about irradiation
of the high-gain register to investigate any kind of unexpected
catastrophic failures.

\section{EMCCD BEGINNING OF LIFE PERFORMANCE}
\label{sec:tests}  % \label{} allows reference to this section

\subsection{EMCCD Camera and Test Conditions}

BOL performance of a ``science-grade'' CCD201, e2v's highest
standard in terms of minimal defects and maximum performance,
was carried out, where the three main criteria assessed
were read noise, dark current, and CIC. We discuss CTI in
Sec. 6. The N{\"u}v{\"u} Cameras EMN2 camera with the CCCP controller\cite{nuvu} was used to perform this
characterization work. The EMN2 is a liquid nitrogen (LN2)
cooled system, allowing deep sensor cooling as low as -105 $\pm$ 0.01$^{\circ}$ C (168.15 K) under calibrated gain. Crucially, the sensor is
driven by the ``CCD controller for counting photons'', or CCCP,
which facilitates full control over all clocking architecture,
allowing waveform design and modification. The motivation
to procure this camera was therefore a direct result of this functionality, enabling JPL to commence an early investigation
into the desired clocking design for optimal CGI operating
conditions.

BOL testing of the CCD201 was carried out over a range of
temperatures using both the conventional and EM outputs as
well as all combinations of vertical clocking frequencies and
horizontal read out rates. A self-pressurized 160-l LN2 dewar,
vacuum-jacketed transfer line and bayonet were used to maintain
the desired temperatures throughout the testing. The device
was run in IMO for maximum dark current performance, and the
serial register was run in its standard NIMO in order to minimize
CIC since EM-generated CIC will dominate in the high-gain
register. The serial register is also flushed in order to clean
residual charge that may have been accumulated. This process
will generate CIC in the leading N pixels; however, typically
read noise will dominate at these levels. For this application
under high multiplication gain, we sought to fully suppress
any CIC component and therefore the next clocking sequence
ignores the first N pixels prior to read out. Initial device bias
conditions as outlined in the CCD201 specifications sheet\cite{e2v}
were used. These were modified thereafter to produce the highest
S/N performance by reducing the read noise and minimizing
CTI within the multiplication register. The high voltage clock
(R$\phi$2HV) voltage was adjusted where necessary in order to
change the multiplication gain applied. We used multiplication
gains of up to $\times$1000 for dark current measurements with the
EM register and for all CIC measurements. This reduced the
effective readout noise to sufficiently low levels so that CIC
and other spurious noise could be detected and assessed. It
also allowed for much shorter integration times for dark current
frames. All dark current data were taken under zero illumination
conditions

\begin{table}
\caption{Measurements of native readout noise (unity gain) of the
CCD201 EMCCD using the N{\"u}V{\"u} EMN2 camera and CCCP controller.
We note that column 5 denotes what gain factor is required based
on the read noise calculation in column 3 in order to meet the
WFIRST-CGI effective read noise requirement of 0.2 e$^{-}$ rms. $k-$gain
is the conversion gain.}
\begin{center}
\begin{tabular}{ccccc}
\hline 
Amplifier & Horiz. Frequency  & Read Noise  & $k-$gain & Gain Factor\\
 &(MHz) &(e$^{-}$) & (e$^{-}$ DN$^{-1}$) & [$\sim$0.2 e$^{-}$ rms] \\
\hline \hline
Conventional & 0.1 & 3.84 & 3.30&...\\
Conventional & 1 & 9.24& 3.27&...\\
Conventional & 3.33 &12.57&3.31&...\\
High gain (EM) & 1 & 33.66&15.30&$\sim$75\\
High gain (EM) & 5 & 75.96&18.99&$\sim$90\\
High gain (EM) & 10 & 89.78&17.82&$\sim$450\\
High gain (EM) & 20 & 253.42&24.13&$\sim$1220\\

\hline
\end{tabular}
\end{center}
\label{table:RN}
\end{table}

\subsection{Read Noise and Conversion Gain}\label{subsec:RN}

The read noise (e$^{-}$ pix$^{-1}$) of a detector is generated during the
readout process, where there is noise associated with the conversion
of charge to an electric impulse as charge packets are
moved through a MOSFET transistor. This noise is limited
by on-chip amplifier noise but can also be attributed to other
sources of noise that are independent of the incoming signal.
Amplifier noise has many components, including Johnson,
Reset, White, Flicker (1/f), Shot, Contact and Popcorn (see Janesick\cite{janesick01} for more information). During the output process,
photoelectrons are converted to an analog signal and subsequently
into a digital number (DN). This conversion is
described by the conversion gain factor, measured in e$^{-}$ DN$^{-1}$.

The native read noise (where no EM gain was applied)
and conversion gain (hereafter k-gain) for the CCD201 were
measured from a photon transfer curve by using standard
techniques,\cite{janesick07} for each output amplifier, at different vertical
frequencies and horizontal readout rates. All data used for
this measurement were bias subtracted, which removes any gradients
or fixed pattern noise since these cumulative effects will
increase any computed standard deviation in raw images. This is
important when assessing any overscan regions, which we
used as additional confirmation of the native read noise. We
show the measured values of read noise and conversion gain
in Table 7, where column 5 denotes the gain factor that is
required, based on the native read noise calculations from the
CCD201, in order to achieve effective read noise performance
of $\sim$0.2$^{-}$ rms, as currently stipulated by the WFIRST-CGI
detector requirements.

\begin{figure}
   \centering
   \hspace{2em}
   \includegraphics[width=15cm]{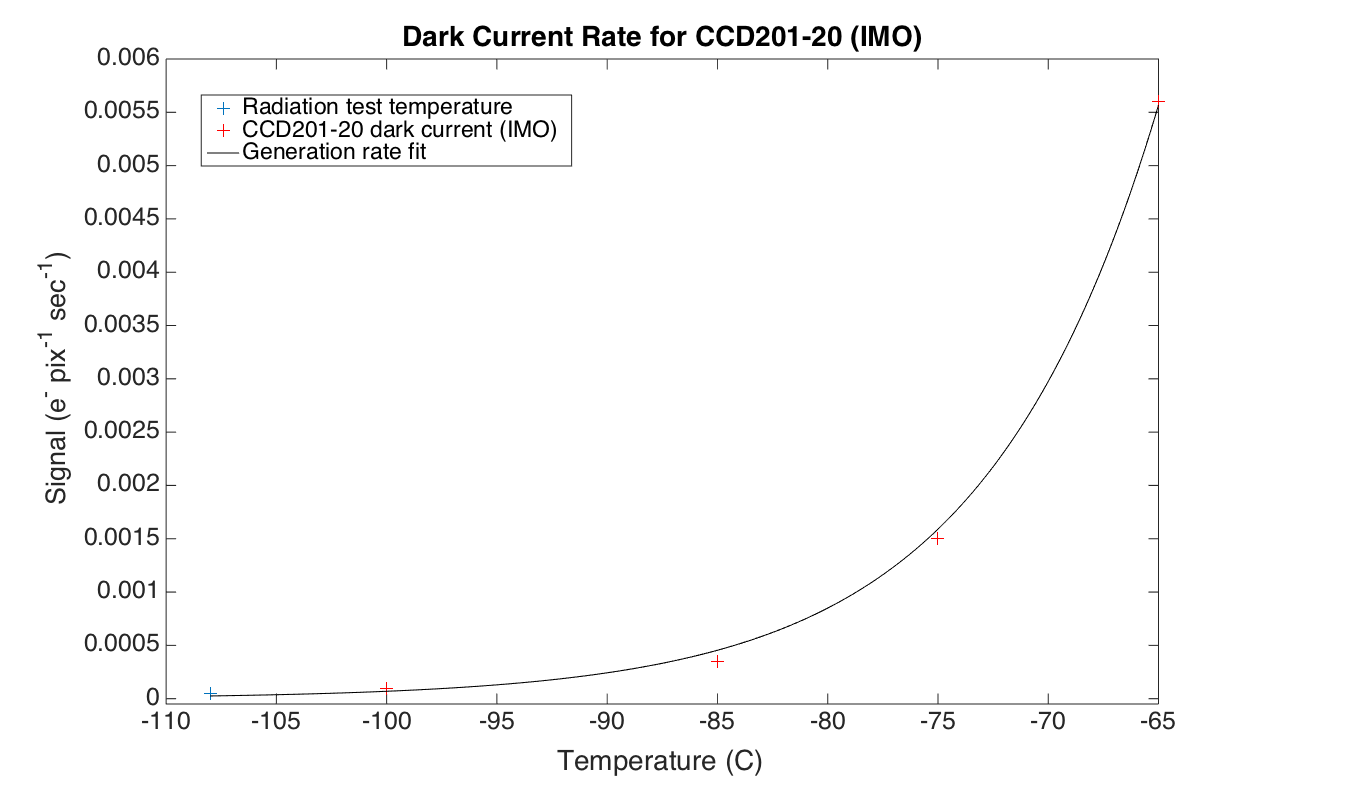}
\caption{Plot of signal as a function of temperature as measured from the CCD201 using the EMN2. For
this characterization, dark current was measured over temperatures from -105$^{\circ}$ C to -65$^{\circ}$ C. We used the
EM amplifier with a gain of 1000 to sufficiently suppress the read noise. The calibrated gain range of
the EMN2 only allows gain to be employed from -65$^{\circ}$ C to -105$^{\circ}$ C. We also include the dark current
datapoint at -108$^{\circ}$ C, which was measured from one of the CCD201s prior to irradiation at the PSI facility,
in Switzerland, which is shown in the bottom left corner as a blue cross.}\label{fig:IDK}
\end{figure}

\subsection{Dark Current}\label{IDK}

Dark current is caused by thermal generation of minority carriers
and is common to all semiconductor devices. Although
running a device in IMO or NIMO can cause large differences
in dark current, ultimately the only solution to suppressing dark
current is achieved by cooling the device. The three main sites in
a device that contribute the most to dark current are bulk regions
below the potential well and channel stops, depleted material in
the potential well, and Si--SO2 interface state contribution from
the front- and backside of thinned CCDs, like the CCD201. See
Janesick\cite{janesick01} for more details. Because we run the CCD201 in IMO, the surface dark current component is greatly reduced,
and dark current is dominated by generation in bulk silicon.
We sought to measure dark current over a range of temperatures
in order to identify a threshold temperature that meets that CGI
dark current requirement.

Dark current data were taken over a range of  -105$^{\circ}$ C to -65$^{\circ}$ C (168.15 K -- 208.15 K) using the EM amplifier under a
multiplication gain of 1000. Under such high amplification,
the read noise contribution becomes negligible, which allows
dark current to be measured over integration times of minutes.
We obtained data for 13 different integration times from 0 to
300 s under zero illumination conditions. Before dark current
assessment, these data were bias subtracted and all CR contributions
were removed. CRs were removed as outlined in
Sec. 4.3. We neglected the outer 100 pixels in both X and Y
in order to avoid edge effects.

The WFIRST-CGI dark current requirement, which is based
on the planetary models described in earlier sections, is
5 $\times$ 10$^{-4}$ e$^{-}$ pix$^{-1}$ sec$^{-1}$, which reflects a dark current at -85$^{\circ}$ C,
see Fig. 10. Since the projected CGI observation time is currently
of order 100--300 s, it is vital that dark current is minimized
as much as possible. Maximum cooling is the first
obvious solution such that thermal excitation is minimized;
however, it has been shown that for temperatures less than
-85$^{\circ}$ C, CTI can occur under high-gain conditions, possibly
as a result of charge clouds interfacing with surface states.
This might cause smearing of CRs or other charge packets
with sufficient energy. Work is ongoing at JPL that is investigating
the tradeoff between dark current and CTI performance for
the WFIRST-CGI. However, the CCD201 passes the dark current
requirement for all temperatures less than -85$^{\circ}$ C. This performance
will degrade as a function of radiation exposure in
orbit; we investigate these effects and provide a discussion in
Sec. 6.

\subsection{Total Spurious Contribution: Clock Induced Charge and Other}\label{sec:CIC}

CIC is a source of spurious generation in CCDs that occurs on
the rising edge of a clock when a phase comes out of inversion.
Therefore, by running a CCD in IMO, surface dark current is
greatly suppressed at the expense of higher CIC.\cite{janesick01} When a
clock pushes the surface into inversion, holes from the channel
stop migrate to the surface but some get trapped in the Si--SO2
interface; however, upon subsequent clocking of the surface
out of inversion, holes are accelerated from the surface to the
channel stops, potentially producing CIC by impact ionization.
The negative-to-positive potential switch under the electrode
supplies sufficient energy for this process to generate and
collect charge in the potential well, also known as CIC. This
generation is a function of the number of transfers, but at a
lower level, the clock amplitude, the clock swing, and the resolution
of the clock edges all contribute to greater CIC. Dark
current and CIC are the most dominant source of detector
noise in the WFIRST-CGI environment, and as a result, strict
requirements have been placed on the detector in order to
achieve $\leq$1.8 $\times$ 10$^{-3}$ e$^{-}$ pix$^{-1}$ frame$^{-1}$ of CIC. This requirement
assumes some margin since modification of the sensor's clocking
architecture may be required based on as yet unknown
CTI, resulting in a trade-off between CIC and CTI. It can be
difficult to disentangle CIC and dark current, in addition to
other sources of noise in a detector. One way to assess CIC is
to take data at zero integration timescales at low temperatures
(e.g., -85$^{\circ}$ C), thereby minimizing the expected thermal excitation
by an order of magnitude with respect to CIC per given
transfer. Therefore, for this component of this work, we have
designed a test that yields low enough dark current signal
so that CIC and other spurious charge can be measured.
Similarly, we employ EM gain to sufficiently suppress readout
noise levels for all horizontal readout rates. We could not calculate
CIC for the conventional amplifier, since any measurement
is inherently read noise dominated.

\begin{table}
\caption{Measurements of total spurious generation (CIC and other)
for the CCD201 EMCCD using the N{\"u}v{\"u} EMN2 camera and CCCP
controller. We note that V$_{SS}$ is the substrate voltage. The CIC quoted
here is the total signal from the image section and the high-gain
register.}
\begin{center}
\begin{tabular}{cccccc}
\hline 
Amplifier & Horiz. Rate  & Vert. Freq.  & EM Gain & V$_{SS}$ & Total Signal \\
 &(MHz) &(MHz) & & (V) & (e$^{-}$ pix$^{-1}$ frame$^{-1}$)  \\
\hline \hline
High gain (EM)  & 1 & 1 & 1000 &4.5004& 1.05 $\times$ 10$^{-2}$\\
High gain (EM)  & 5 & 1&1000&4.5004&3.20 $\times$ 10$^{-3}$\\
High gain (EM)  & 10 &1&1000&4.5004&1.25 $\times$ 10$^{-3}$\\
High gain (EM) & 10 & 0.2&1000&4.5004&8.75 $\times$ 10$^{-3}$\\
%High Gain (EM) & 10 & 0.8&1000&4.5004&8.80 $\times$ 10$^{-3}$\\
High gain (EM) & 20 & 1&1000&4.5004&2.20 $\times$ 10$^{-3}$\\

\hline
\end{tabular}
\end{center}
\label{table:CIC}
\end{table}

Zero integration, zero illumination frames taken at -85$^{\circ}$ C,
were obtained using the EM amplifier for each horizontal readout
rate as well as different combinations of vertical clock
frequencies, where a gain of 1000 was used in all cases. An output
histogram of 20 frames per given acquisition mode combination
was produced, and a Gaussian fit was applied to the read
noise section of this plot, as shown in Fig. 11. We applied a 5.5$\sigma$
threshold, indicated by the red dashed vertical line, to the mean
of the Gaussian plot, which represents the mean of the read
noise. This threshold was best representative of the point at
which spurious signal was getting amplified by the high-gain
register. We then integrated under the curve to the right of
this threshold and calculated the total signal, per pixel, per
image. Based on a temperature of -85$^{\circ}$ C, we removed any
expected dark current contribution as measured in Sec. 5.3. The
result was a measure of all spurious generation in the system
for a given mode, including CIC and other. We show these measurements
in Table 8. We note, however, that measuring CIC
using this method can sometimes be misleading if a system's CTI in the serial direction has not been characterized well.
We show the measured CIC signal decreasing with increasing
serial clock speed where this rate is roughly inversely proportional
to clock frequency. However, there is a small increase
from 10 to 20 MHz that does not follow the trend. If CTI
effects are present for the 20-MHz timing, charge packets
might spill into a pixel that follows another, or beyond, and will increase the number of pixels that exceed the 5.5$\sigma$ threshold
outlined previously. We show another way of measuring
CIC in Sec. 6.4.3.

\begin{figure}
   \centering
   \includegraphics[width=14cm]{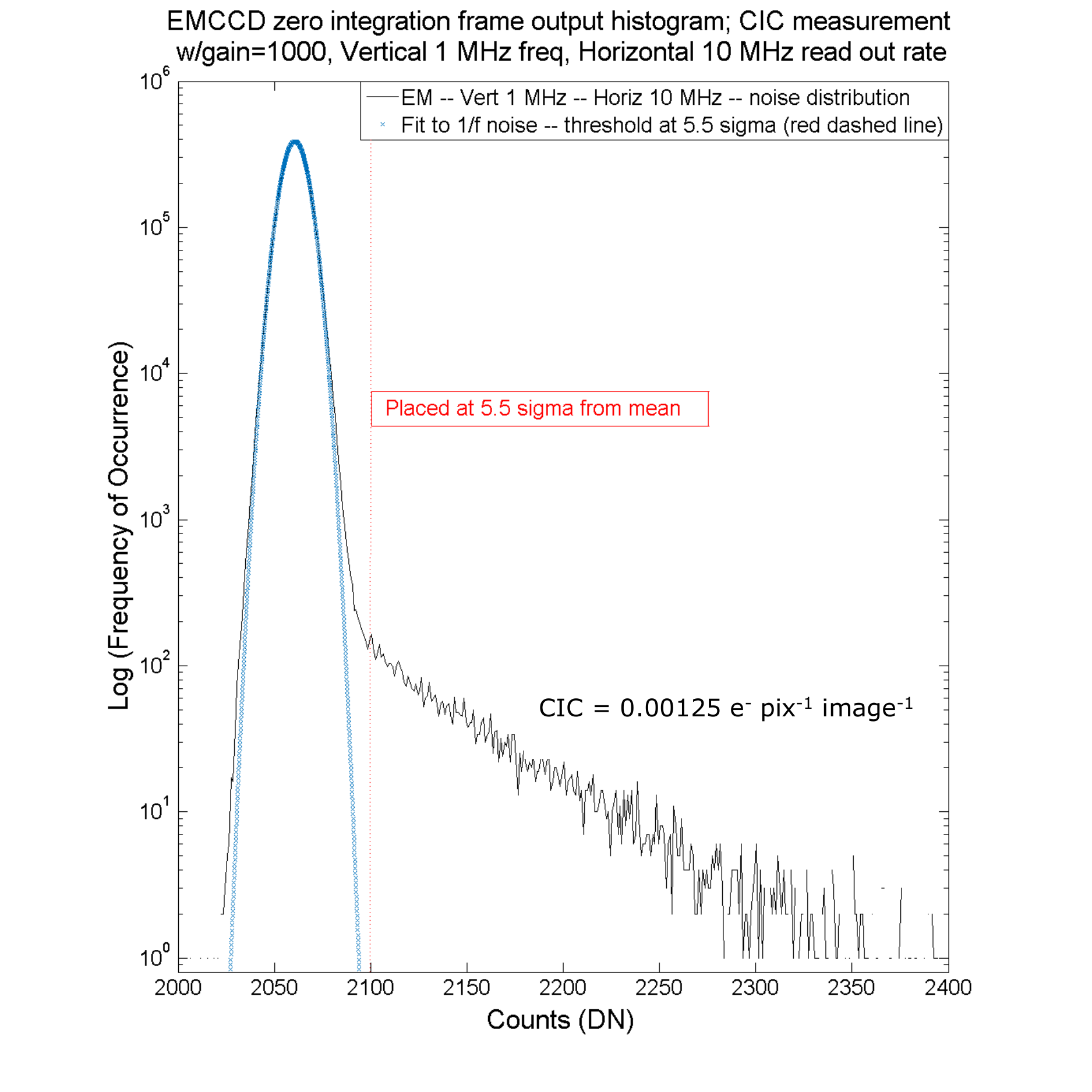}
\caption{Histogram of 20 EMCCD frames, taken at a gain of 1000. The
Gaussian function to the left is the read out noise (primarily 1/f ) with
small amounts of CIC and dark current. The data points to the right of
the 5.5$\sigma$ are the total contribution of CIC, dark current and other spuriously
generated charge. The 5.5$\sigma$ threshold was applied to the mean
of the blue Gaussian fit to the read out noise. When calculating CIC,
we integrate under this curve, and scale by the total number of pixels
per given image, and remove the dark current for this temperature.
The result is charge generated from CIC and other spurious sources.}\label{fig:CIC}
\end{figure}

Finally, we highlight that JPL is currently conducting an
investigation as to why the total signal, as measured in these
data, appears to be higher for lower readout rates, where alternate
clocking techniques are being considered. Since the CIC
mechanism is believed to be due to holes escaping from traps at the surface and producing electrons by impact ionization
given sufficient energy, the longer there is a high potential
difference between two phases, the higher the probability that
a hole will escape and be subjected to a higher field. It may
be this time dependence that results in faster clocks producing
less CIC. Importantly, as noted previously, any decreasing serial
CTI might contribute to apparent higher CIC levels; therefore,
CTI measurements for all clocking implementations will also be
measured. These considerations become important for flight
electronics, since higher readout rates are inherently more difficult
to accommodate for flight, based on requisite components
that are highly temperature and radiation sensitive; these components
are unique to a design that facilitates faster clocking.
This work is ongoing.

\subsection{Reducing the Noise Contribution}

Since the total spurious noise in an EMCCD, which we define to
be the contribution from dark current and CIC, can be amplified
by the EM register, it is the dominant source of noise since readout
noise is suppressed under such amplification. As discussed
previously, thermal excitation resulting in dark current can be
greatly reduced by deep sensor cooling in addition to running
the device in IMO. However, since the CIC can vary strongly as
a function of clock swing, clock rise time as well as the device
bias conditions (in addition to waveshapes, as reported by
Daigle et al.\cite{daigle08}), it is thus heavily dependent on the clocking
process and how these parameters are designed in order to transfer
charge.\cite{janesick01} Indeed, careful consideration of a device's clocking
design also minimizes the CTI. Here, we show the different
sources of noise inherent to an EMCCD, which must be considered
in all performance modeling of such a device, and highlight
these differences when compared to conventional devices.
Additionally, we outline the different operating modes and the
importance of precision clock timing in order to reduce CTI for
the WFIRST-CGI application.

\subsubsection{Noise in EMCCDs}

Evaluating the S/N in an EMCCD is much like that of a conventional
CCD; however, since there is a gain factor applied during
the EMCCD amplification process, the equivalent readout noise,
$eff_{RN}$, is reduced by an amount $R/G$, where $R$ is number of electrons
and $G$ is the amplification gain. However, as previously
outlined in Sec. 2, there is an added variance in the signal,
the ENF, based on the stochastic nature of the process.
Therefore, when calculating the S4N for an EMCCD, the
read noise becomes negligible but the ENF and multiplication
gain must be included, in addition to other standard parameters,
such as the QE and the total signal in photoelectrons, $S$, where $S=P \cdot QE \cdot t$. The terms $P$ and $t$ are the incident photons and
integration time, respectively. The principle sources of noise are
found via the root of sum of squares law and the S/N equation is
thusly calculated as shown (adapted from McLean et al.\cite{mclean08}):

\begin{equation}
S/N=\frac{S}{\sqrt{2S + (\frac{R}{G})^{2}}},
\end{equation}
where $S$ becomes $2S$ in the denominator representing the noise, because of the ENF adds to the photon shot noise where the uncertainty now becomes $\sqrt{2S}$ as opposed to $\sqrt{S}$ (ENF$=1.41^{2}\sim2$).  Most importantly, although the read noise gets reduced by $R/G$, the pixel charge handling capacity will also be reduced by this same factor.  Although the CGI application will produce an expected photon flux in flight of order 2 $\times$ 10$^{-2}$ photons pix$^{-1}$ sec$^{-1}$ in science acquisition mode, and 5 $\times$ 10$^{-4}$ photons pix$^{-1}$ sec$^{-1}$ in other acquisition modes, the pixel's charge capacity must be considered in order to avoid unwanted saturation in the high gain register should a bright source spuriously contaminate the field.  Other factors such as charge persistence must be considered in such scenarios.  

\subsubsection{Persistence}

Persistence can occur when there is residual charge trapped in
the surface or bulk regions of an EMCCD, and typically occurs
if the device has been over-exposed to light such that the pixel
charge capacity is exceeded. There are two main types of persistence,
also known as ``residual image'': (1) residual surface
image (RSI) and (2) residual bulk image (RBI).

RSI can cause signal charge to get trapped in the Si--SO2
interface, which may persist at these positions and subsequently
get released during a readout. Charge trap lifetimes can be long
and so persistence can be seen for hours or weeks after it
occurs.\cite{janesick01} However, the use of IMO in the CCD201 eliminates
surface trapped charge immediately by flooding the Si--SiO2
interface with holes; therefore, the RSI component is eliminated.
In parts of the CCD that are never inverted, such as the high-gain
register, RSI is a potential issue. However, the high-gain register
pixel well capacity is much larger than the image/store pixel
well capacity, so it is unlikely that charge will ever reach the
surface under normal, unbinned operation. In order to investigate
potential damage or persistence effects, testing is currently
ongoing at JPL, where EMCCDs are being subjected to the
highest predicted signal levels from a CGI observing mode.
In the event that the high-gain register becomes saturated, it
has a built-in mechanism to prevent overloading from occurring.

Vertical antiblooming in a device can negate the effects
of RBI; however, there is no vertical antiblooming in the
CCD201. With back thinning, RBI is reduced compared to
front-illuminated CCDs because the epibulk interface is eliminated
by the thinning process. However, a small amount of
RBI remains, from unknown sources.

\subsubsection{Mode of operation and cooling}\label{sec:IMOvsNIMO}

Since the CCD201 contains two boron barrier implants under
two of its phases, it can be clocked in IMO. In this mode,
the substrate voltage (V$_{SS}$) is raised to a level higher than
that of the potential at the semiconductor--insulator interface
beneath the gate electrodes. Once this condition has been fulfilled,
holes migrate from the channel stops of the device to
the semiconductor--insulator interface and populate the traps
that are present due to the atomic mismatch between Si and
Si--O2. Since the traps are populations with holes, dark current
generation from this region is significantly reduced. Therefore,
IMO is favored over NIMO, and by using IMO for the CGI
application, which includes individual integration times of
order 100's of seconds and accumulated total integration
times of several 1000 s, the dark current requirement of
5 $\times$ 10$^{-4}$ e$^{-}$ pix$^{-1}$ sec$^{-1}$ can be achieved at and below temperatures
of -85$^{\circ}$ C. In flight, we expect the sensor to be passively
cooled with active warming.

Although running the device in NIMO can greatly reduce
CIC, as stated previously, dark current accumulation is likely too high in this mode. We would need to use temperatures of
approximately −115$^{\circ}$ C and below to sufficiently suppress the
dark current in NIMO; however, under high gain, we are still
investigating if temperatures in this range will cause increased
CTI. In the following section, we describe recent work that has
been carried out at the manufacturing system level that offers
lower parallel CIC.

\subsubsection{Clock induced charge minimization}\label{CIC-minimize}

CIC is highly dependent on the clocking process,\cite{janesick01} and therefore
at the camera system level, custom clocking parameters, including
careful consideration of V$_{SS}$, clock swing, and clock amplitude,
can be considered to minimize these effects. We note that
other groups claim that wave shaping can also play an important
role in CIC minimization\cite{daigle08}. These efforts are ongoing at JPL to
achieve similar spurious noise contribution at lower horizontal
read out rates. Since CIC levels are affected by clock amplitude,
and since the CIC will rise exponentially with the clock amplitude\cite{maes90}
under EM gain, it follows that lower clock amplitudes
will produce lower CIC. There is a minimum voltage that
must be used in order to move charge in the parallel section;
however, as a result of the barrier implants described previously,
a fixed potential is present in the silicon and any charge must
overcome this static field in order to maintain sufficient device
charge transfer. In recent work by Gach et al.\cite{gach14}, they outline a
strategy to reduce the effective implant dose of an EMCCD
(CCD282), consequently lowering the required clock amplitude
to maintain charge transfer. Impact ionization is therefore
reduced in the local electric field. This reduction in implant
dose will also mean a reduction in the FWC of the device; however,
a tradeoff between dose reduction and FWC is dependent
on the application. This modification could also be applied to
the CCD201 and might be considered if CIC levels become
problematic. Additionally, increasing the number of multiplication
elements ($N$) would result in the same given gain factor ($G$)
being achieved by using a lower gate high voltage, since $G \propto N$,
as previously shown in Eq. (1). By producing lower EM gain,
less impact ionization would occur due to lower electric fields
thus reducing CIC. The CCD201 has 604 multiplication elements
in its standard design, and therefore this would require
a further design change.

Another way of reducing parallel IMO CIC to NIMO CIC
levels is via multilevel clocking of the image/store array (see
Murray et al.\cite{murray13} for discussion of such clocking techniques).
During an integration, all of the clocks are held low enough
with respect to V$_{SS}$ so that the image array is fully inverted.
After this integration, but before the image to store transfer,
the low rail of the image/store clocks is raised (or alternatively,
V$_{SS}$ is dropped) enough so that the image/store array is taken out
of inversion. The device is therefore readout in NIMO mode. At
higher temperatures than -108$^{\circ}$ C (165 K), NIMO dark current
may no longer satisfy CGI requirements. Once the CGI operating
temperature has been selected such that the dark current and
CTI are minimized under high gain, this optimization can be
considered further.

\subsubsection{Clocking considerations: charge transfer inefficiency}

CTE is defined as the efficiency of a device to transfer charge
from one pixel to the next and is an extremely important
parameter that affects a CCD's overall performance, where
CTE $=$ 1 -- CTI. CTI can vary based on the vertical clocking frequency, wave shape, and horizontal frequency, in addition
to the temperature of the device with high multiplication gain
(and because of radiation damage, see Sec. 6). Indeed, by designing
custom clock signals in an attempt to minimize other sources
of noise, e.g., CIC, the CTI can be adversely effected, see, e.g.,
Murray et al.\cite{murray13} and Hall et al.\cite{hall14d} The data presented in this section
was produced by using the CCCP version 3 controller,\cite{daigle14} which
has been designed to minimize both vertical and horizontal CTI
for photon counting applications that can yield CTI of order
$\sim$1 $\times$ 10$^{-5}$. This controller has a core operating frequency of
240MHz and produces 4.16-ns time resolution for all analog signals.
Digital CDS can process up to 2048 samples per pixel. It
generates 14 analog clocks ($\pm$15 V) where two high-voltage
clocks, with resonant frequencies of 10 and 20 MHz with 68-
ps resolution, are used. The CCCP therefore generates clock
and bias signals with precision timing and produces low dark current
(run in IMO) and low CIC (clocking architecture), while
maintaining sufficient charge transfer

As discussed in Sec. 4, the CCD201 will undergo radiation
damage in the WFIRST orbit, and therefore a clocking design
that is sufficient to deliver the required CTI performance at BOL
might not be applicable as damage accrues over the mission lifetime.
Clocking strategies will therefore be required to identify
damaged sites on the sensor, as well as additional techniques to
mitigate this damage and/or maintain charge transfer quality
under irradiated conditions. The performance of devices, as
shown in, e.g., Daigle et al.,\cite{daigle08, daigle14} Gach et al.,\cite{gach14} and the strategies
used to obtain the resulting S/N, have yielded extremely promising
results for devices at BOL. These strategies and techniques
are being developed further at JPL specifically for irradiated
CCD201s to meet the CGI requirements at EOL. Details of
this work will be published in a separate document at a later
time. In Sec. 6, we outline some preliminary clocking techniques
that were designed as a tool to investigate displacement
damage in the CCD201.

\subsubsection{Manufacturer design modifications}\label{design-mod}

After selection of the CCD201 for the WFIRST-CGI, in addition
to the performance differences between a standard thickness silicon
device and a deep depleted device (see Sec. 1.2), we also
considered design modifications at the device processing level
that could yield higher S/N performance. Two main modifications
were considered: (1) reducing the channel width of the
device and (2) reducing the device's barrier implant dose.

A reduction in the channel width would reduce the charge
store volume of the image and store sections. This would have
the effect of reducing the amount of trapping sites that a charge
cloud could interact with during readout, thus decreasing the
CTI. We note that discussions with e2v have suggested that
this is considered to be a nontrivial design change. Reducing
the implant dose under phases 1 and 3 of the CCD201
would yield negligible levels of CIC in the parallel section;\cite{gach14}
however, a tradeoff between implant does reduction and
a similar reduction in FWC must be considered. Importantly,
by completely removing the implant, IMO would no longer be
possible, which would affect the dark current.

Since the standard off-the-shelf CCD201 has higher maturity
and lower fabrication risk than a custom design, we decided to
implement no manufacturer design modifications at this time
and baselined the standard device for the WFIRST-CGI.
Although it is not desirable to propose modifications to a sensor
that has been selected based on its maturity, if EOL performance as indicated by the radiation phases I and II test plans does not
meet the CGI requirements to the point that the CCD201 can no
longer provide sufficient planetary yield for the mission, then
these modifications will be reconsidered as well as substitution
of other flight-rated conventional CCDs.

\section{RADIATON STUDY OF CCD201-20}
\label{sec:rad-study}  % \label{} allows reference to this section

\subsection{Test Plan and System Calibration}

\subsubsection{Phase I irradiation plan}

\begin{figure}
   \centering
   \includegraphics[width=15cm]{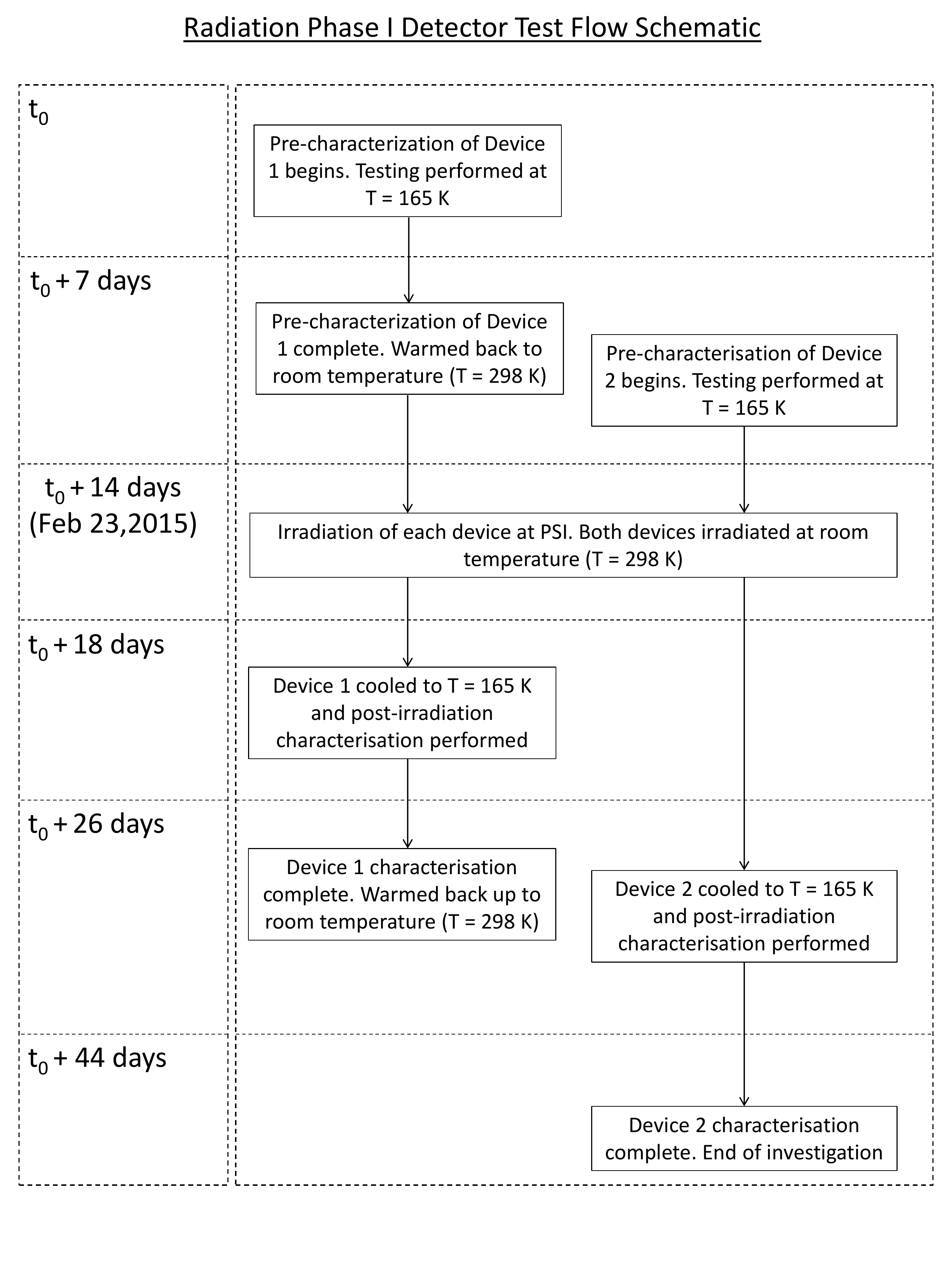}
\caption{Radiation phase I detector test flow diagram. The left column of the schematic shows the
beginning of the test, t$_{0}$, where we include each major step in the program until the end of the test,
 t$_{0}$ + 44 days. The right column shows each subsequent stage of the program for each device, including
when the radiation occurred as well as details of timings, cooling, and temperatures.}\label{fig:CEI-test-flow-diagram}
\end{figure}

In Sec. 1, we outlined a two-phase radiation study to investigate
DDD damage on the performance of the CCD201. In this
section, we present the results of phase I, where an ambient
temperature, unbiased, irradiation was carried out at the proton
irradiation facility (PIF), at the Paul Scherrer Institute (PSI), in
Switzerland, on February 23, 2015, and postirradiated analysis
was carried out at CEI, in the United Kingdom. This study was designed to include full sensor characterization of two engineering
grade CCD201 devices at ambient temperature. It
should be noted that although these devices are considered engineering
grade, we consider the measured performance in this
study postirradiation as representative of what scientific grade
sensors will experience, since this study sought to investigate a
degradation factor on performance for each characterization
parameter. These parameters include multiplication gain, dark
current, CIC, high flux parallel and serial CTI as well as low
flux parallel and serial CTI, before and after a radiation dose of
2.5 $\times$ 10$^{9}$ protons cm$^{-2}$ (10 MeV equivalent). This dose reflects
the EOL dose (6 years in L2) for a 10-mm tantalum shield. We
show a detector test flow diagram in Fig. 12. Phase II commenced
in June 2015 and covered the full range of doses for
all of the shielding materials and thicknesses shown in Fig. 6,
where a custom apparatus was designed and built to maintain
device cryogenic temperatures throughput the study.

We selected irradiated regions of each device thereby ensuring
a control region for both parallel and serial registers--this allowed the independent and combined effects of radiation-induced
displacement damage on parallel and serial CTI to
be investigated in a meaningful way that prevented contamination.
A schematic of this control scheme is shown in Fig. 13.
Monte Carlo simulations were performed using the Stopping
Range of Ions in Matter package (SRIM).\cite{ziegler10} These simulations
provided the maximum penetration depth of the beam as it
passes through stainless steel (the element used for shielding),
which was $\sim$9 mm, thus a thickness of 14 mm was used for
redundancy.

For device 1, the shielding profile was designed such that the
degradation of parallel performance could be evaluated where
serial and high-gain registers were protected. Additionally,
the right side of the parallel section was also protected and
used as a control region, where results from this section were
expected to be identical to those prior to irradiation. The shielding
profile in device 2 allowed the effects of radiation damage to
be assessed on both the serial and parallel sections, as well as
these sections combined. In this way, charge getting transferred
through the right side of device 2 should get affected in the serial
direction only. Note that in device 2, a small region of the parallel
section in the control region (right side) was irradiated;
this is because the serial register is small when compared to
the store and image sections, and therefore irradiating a small
region ($\sim$300 pixels) of the parallel section ensured that the
serial register was fully irradiated.

\begin{figure}
   \centering
   \includegraphics[width=12cm]{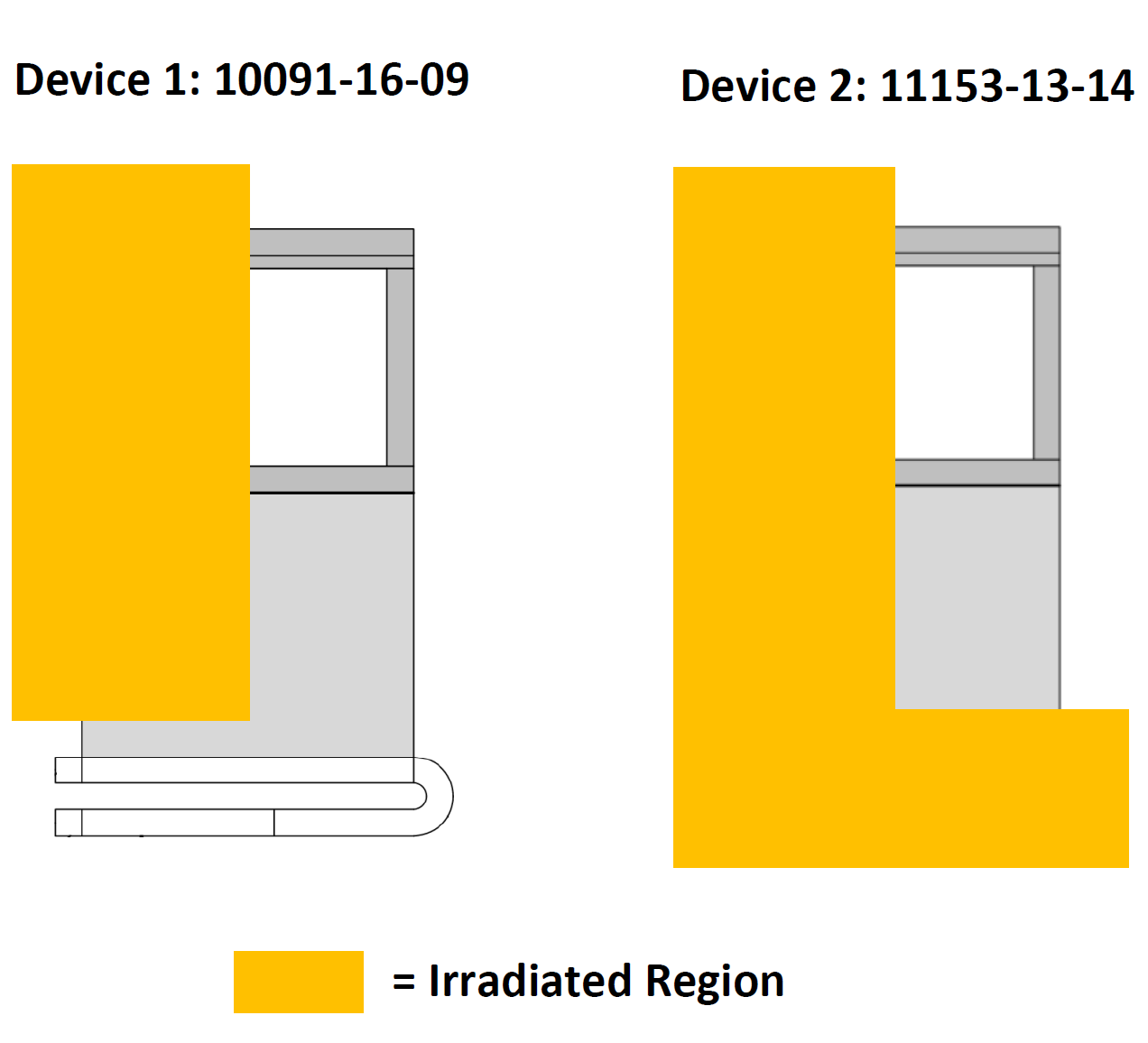}
\caption{Irradiation pattern for each CCD201 device. We note that the
relative size of the serial and multiplication registers is not to scale in
this image. For device 1, the shielding covered the first $\sim$300 pixels of
the store section, as shown, in order to prevent any effects of secondary
nuclear interactions in the proximity of the shielded edge.}\label{fig:CEI-control}
\end{figure}

\subsubsection{Device parameters and calibration}

\textit{CCD controller.} All pre- and postirradiation
tests were carried out at -108$^{\circ}$ C (165 K) with a
frame integration time of 100 s (unless otherwise stated) and
the devices were run in IMO. The CCD was operated using
generic drive electronics for test and characterization from
XCAM Ltd.,\cite{xcam} consisting of a 19$^{\prime\prime}$ inch rack-based controller,
together with proximity electronics on a headboard PCB
local to the CCD, providing bias filtering and low noise preamplification.
The control electronics used for testing were standard
and thus were not fully optimized to match those of the
EMN2 described earlier in Sec. 5. We note that the use of this controller does not invalidate this test, since we sought
to identify performance degradation of the CCD201 only--
this does not include controller or other proximity electronics.
While some parameters will change based on the camera controller
used (e.g., readout rate), the trends exhibited due to radiation
damage in this study should be independent of the CCD
controller. Part of the study strategy was to keep pre- and postirradiation
controller operating conditions identical. We felt that
it was important to initially separate the detector and the controller
so that we could interpret results in the context of the
EMCCD only. The EMCCD flight electronics will be designed
for performance in the relevant environment.\\

\textit{Calibration.} The XCAM setup enabled rapid investigation
of the pre- and postperformance of the device. We used the
default readout modes for clocking through the LS amplifier
and bias signals as recommended by e2v\cite{e2v} were used with
the device image and store sections being operated in two phase
mode. Some minor modifications were made based on
precharacterization tests in order to minimize the readout
noise and a read out pixel rate of 700 kHz and parallel line transfer
time of 6 $\mu$s per phase were employed. We used the drive
system with the option of XCAM's two-channel CDS card,
where with the combination of headboard gain and CCD output
node responsivity, the system noise for EM gain $=$ 1$\times$ is dominated
by the noise on the 14-bit ADC (rather than the CCD
itself). However, operating with a maximum amplification
gain of $\sim$200, we were able to routinely operate with an effective
readout noise of 0.8 e$^{-}$ rms. This low effective noise was important
in order to minimize the CTI (from both small optical
signals and $^{55}$Fe x-rays) within the multiplication register. Once
optimized, we kept these bias conditions consistent for pre- and
post-analyses. Similarly, the same R$\phi$2HV (this voltage controls
the multiplication gain) was consistent in both scenarios.

The amplifier responsivity and conversion gain were measured
in order to calculate the system calibration, $DN/e^{-}$, as
follows:

\begin{equation}
\frac{DN}{e^{-}}=\frac{V}{e^{-}} \cdot \frac{DN}{V},
\end{equation}
where $V/e^{-}$ is the amplifier responsivity and $DN/e^{-}$ is the
k-gain. We note that $DN/e^{-}$ is the system calibration$^{-1}$. This
calculation was necessary since it provided the signal level
as measured from the phase I device and system electronics
and was essential for measurements later. Obtaining these
parameters prior to irradiation was also a good indication of
a change to the output electronics after the radiation tests
were carried out. An $^{55}$Fe x-ray source was used for this component
of calibration.

\subsection{Proton Fluence and Beam Profile}

In order to achieve a proton fluence of 2.5 $\times$ 10$^{9}$ protons cm$^{-2}$
(10 MeV equivalent), we used the PIF at PSI, Switzerland,
which was been used by ESA for major testing of spacecraft
components. The key features of the facility that allowed this
radiation dose to be achieved are as follows:

\begin{description}

	\item[$\bullet$ Lowest initial proton energy:] 74 MeV.
	\item[$\bullet$ Beam profile:] Gaussian with typical FWHM of 10 cm.
	\item[$\bullet$ Neutron background:] $<10^{-4}$ neutrons proton$^{-1}$ cm$^{-2}$.

\end{description}

As indicated previously, since the lowest primary beam
energy at PSI is 74 MeV, we needed to extrapolate the final
dose required at 74 MeV to deposit equivalent energy levels
into the CCD201s, as a 10-MeV beam at the required fluence.
We therefore adopted the Non Ionizing Energy Loss (NIEL) function,
as found in Burke et al.,\cite{burke86} Vanlint et al.,\cite{vanlint87} and Srour et al.:\cite{srour03}

\begin{equation}\label{eq:6}
\mbox{10 MeV NIEL Function}=\frac{1.6}{E_{p}^{0.28}},
\end{equation}

\begin{equation}
\mbox{10 MeV NIEL Function}=0.479,
\end{equation}

\begin{equation}
\mbox{Fluence at 74 MeV}=\frac{2.5 \times 10^{9}}{0.479},
\end{equation}

\begin{equation}\label{eq:9}
\mbox{Fluence at 74 MeV}=5.21 \times 10^{9} \; \mbox{protons cm$^{-2}$}
\end{equation}

Prior to irradiation, the intensity of the PIF beamline was
mapped around the target area to confirm that the beam intensity
across the irradiated regions was consistent with the above calculation
in Eq. (9). A normalized beam intensity difference of
$<$8\% was measured from the center of the CCD201 (where the
image and store sections meet) to the outer edge of the device
(top of image section and bottom of serial section), which we
found to be sufficient for this study.

\subsection{Trap Pumping: Initial Identification of Damage}\label{trappump}

Although the majority of trapping sites in an irradiated device
are a direct result of displacement damage, some traps will be
present as a result of imperfections caused during the manufacturing
process, e.g., poor cleaning between the silicon base and
the first polylayer is a big factor in the number of these so-called
``design traps''. Such sites were therefore assessed prior to and
postirradiation in order to (1) highlight the irradiated regions
(where new traps now exist) and (2) determine the approximate
trap density and distribution in the device.

Standard pocket pumping\cite{janesick01} was employed, which provides
the means of probing trap sites, where pixels were clocked in the forward direction and then subsequently in the opposite
direction. We highlight again that since the CCD201 device
is designed to be operated in a two-phase mode with charge
only flowing forwards, each of the four phases is independently
connected and therefore the device can be operated as a four phase
device, albeit provided that one considers the impact
of the implants on the clocking. To do this, we paired $\phi$1
and $\phi$2 together and $\phi$3 and $\phi$4 together, to push charge forward,
and then paired $\phi$2 and $\phi$3 and $\phi$1 and $\phi$4 together, to push
the charge backward. This technique was then repeated, which
transfers charge between adjacent gates within the same pixel
and thereby captures traps within the signal packet which are
then released into a neighboring pixel when the original signal
packet has moved from the trapping location. As a result of this
process, a pair of bright and dark pixels is now present signaling
a trapping site and can otherwise be known as a ``dipole''. Note
that all of the trap populations will not be present as a result of
the pumping cycle described above. This is because trapping
efficiencies are a function of temperature and also parallel
line transfer timing, and each pumping cycle will only probe
a certain area of the pixel. Furthermore, the pocket pumping
technique used to generate the images as shown in Fig. 14
did not probe 100\% of the pixel, where we estimate $<$50\%
was probed; this is a tentative estimate since it is difficult to
know the effect of fringing fields within a pixel. Work is
ongoing to probe a larger area of the pixel, which is difficult
based on the presence of barrier implants.

\begin{figure}
   \centering
   \includegraphics[width=15cm]{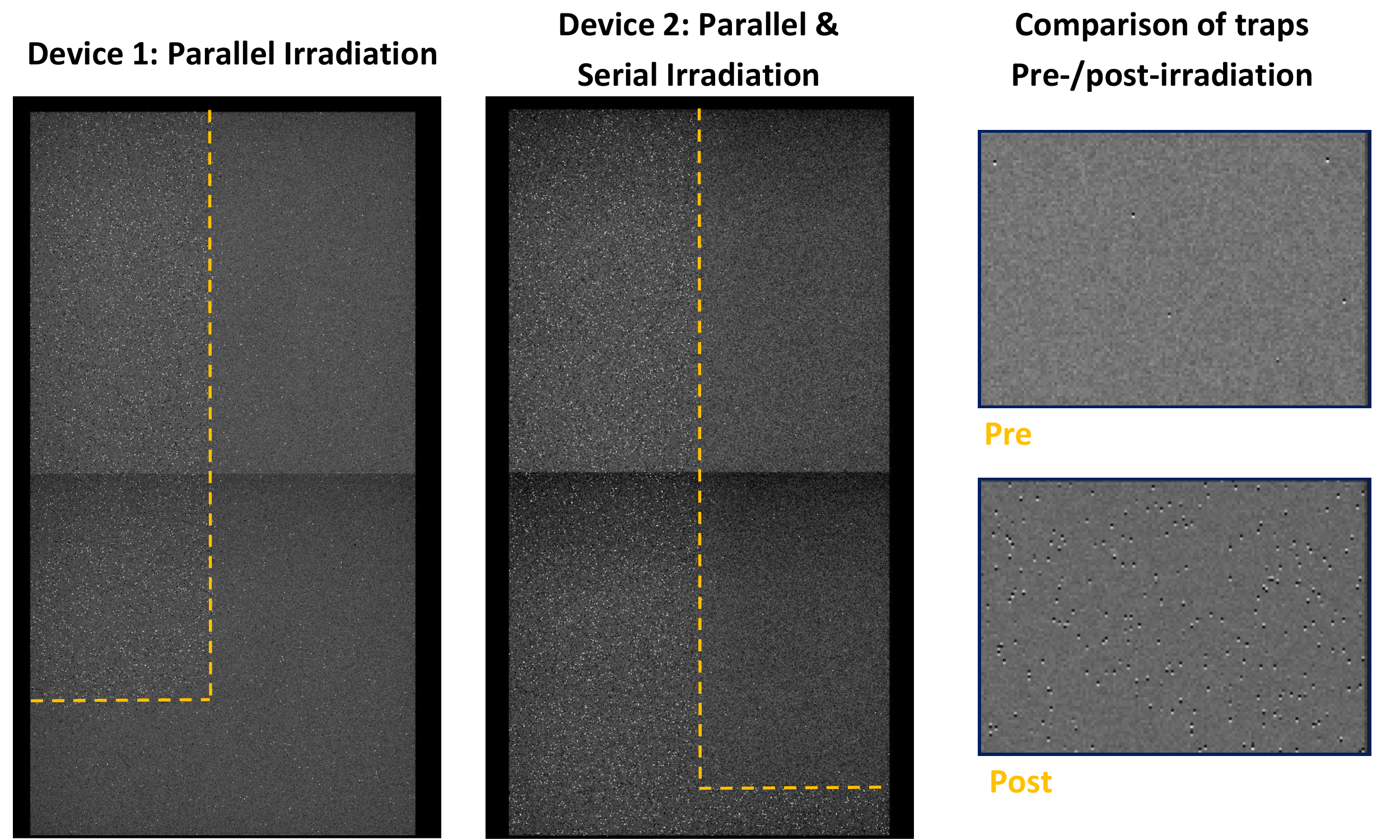}
\caption{Basic pocket pumping that was carried out in order to highlight the unirradiated and irradiated regions of device 1 (left) and device 2 (center). We also show a comparison trapping site (right), highlighting a region of 90 $\times$ 50 pixels in the image area.  In all cases, it is clear that the irradiated regions (within the dashed boxes) contain a much higher density of dipoles.}\label{fig:CEI-DDD}
\end{figure}

We show the results of this analysis in Fig. 14, where new
dipole populations are clearly evident and in greater number as a
result of displacement damage. We cannot offer a \%-increase in
population at this time, since many pocket pumping iterations
are required to probe disparate trapping efficiencies, as outlined
above. We refer the reader to Hall et al. (2014)\cite{hall14a, hall14b, hall14c}, Wood et al (2014)\cite{wood14}, and Murray et al. (2012)\cite{murray12b, murray12a} for in-depth studies of trap pumping for
radiation-damaged devices.

\subsection{Results: Pre- vs. Post-Irradiation}

Here, we present displacement damage radiation results for
the CCD201. This is the first study to be carried out with a CCD201 in assessing performance degradation of this kind.
We can confirm that both CCD201 EMCCD devices were
fully functional after irradiation and experienced the degree
of degradation in multiplication gain, dark current, CIC, and
charge transfer that has been previously reported by other CCD
sensors\cite{srour03} irradiated to a similar dose. We summarize the key
performance degradation below and show comparison results
for each device in Table 9:

\begin{itemize}
  \item Each device was fully functional following irradiation to 2.5 $\times$ 10$^{9}$ protons cm$^{-2}$ (10 MeV equivalent).
  \item No significant change in the multiplication gain was observed (for an unbiased irradiation at ambient temperature).
  \item No significant change in the parallel CIC was observed (for an unbiased irradiation at ambient temperature).
  \item There was an IMO dark current increase by a factor of $\sim$2.
  \item For low signal parallel CTI ($\sim$8 e$^{-}$), there was a degradation by a factor of $\sim$5.  The serial CTI appears to be dominated by the EM register and interaction with surface states for the operating conditions that were used.
  \item Initial results suggest that serial CTI of the EM register has some dependence on the size of the signal being transferred as well as values for $\phi$DC and R$\phi$2HV.
\end{itemize}

\begin{table}
\caption{Summary of pre- and post-irradiation test results for both CCD201 EMCCD devices.  Any degradation factor that is left blank we consider to be inconsistent.}
\begin{center}
\begin{tabular}{cccccc}
\hline
Parameter & Device 1  & Device 1  & Device 2 & Device 2 & Degradation \\
 &Pre-Irradiation &Post-Irradiation & Pre-Irradiation & Post-Irradiation & [factor]\\
\hline \hline \\
Multiplication Gain  &238 $\pm$ 7 & 202 $\pm$ 6 &237 $\pm$ 8 &237 $\pm$8&...\\
\hline \\
Eff. Read Noise  &0.85 $\pm$ 0.04 & 0.98 $\pm$ 0.06 & 0.70 $\pm$ 0.07 & 0.91 $\pm$0.04&$\sim$1.2\\
(e$^{-}$ pix$^{-1}$) & [Par.] & [Par.]  & [Par. + Serial] & [Par. + Serial] &\\
\hline \\
Par. I$_{DK}$ & (7.13 $\pm$ 0.49)  & (1.22 $\pm$ 0.06) & (3.195 $\pm$ 0.50)  & (5.75 $\pm$ 0.44)& $\sim$1.8 \\
 (e$^{-}$ pix$^{-1}$ sec$^{-1}$) & $\times$ 10$^{-5}$ &  $\times$ 10$^{-4}$ & $\times$ 10$^{-5}$ & $\times$ 10$^{-5}$&\\
\hline \\
Par. CIC &  (4.47 $\pm$ 0.24)  & (4.86 $\pm$ 0.26) & (8.80 $\pm$ 0.38)  & (9.90 $\pm$ 0.10) & ...\\
  (e$^{-}$ pix$^{-1}$ fr$^{-1}$)& $\times$ 10$^{-2}$ &  $\times$ 10$^{-2}$ & $\times$ 10$^{-3}$ & $\times$ 10$^{-3}$&\\
\hline \\
$^{55}$Fe Par. CTI &(5.14 $\pm$ 3.90)  & (4.05 $\pm$ 0.54) & (1.23 $\pm$ 0.59)  & (3.07 $\pm$ 0.08)&$\sim$10\\
 & $\times$ 10$^{-6}$ &  $\times$ 10$^{-5}$ & $\times$ 10$^{-6}$ & $\times$ 10$^{-5}$&\\
\hline \\
$^{55}$Fe Serial CTI & (1.05 $\pm$ 1.36)  & (2.04 $\pm$ 2.66) & (1.73 $\pm$ 1.49)  & (1.08 $\pm$ 0.67)&...\\
 (Conv. register) & $\times$ 10$^{-6}$ &  $\times$ 10$^{-6}$ & $\times$ 10$^{-6}$ & $\times$ 10$^{-5}$&\\
\hline \\
$^{55}$Fe Serial CTI &(1.19 $\pm$ 0.01)$^{\dag}$  & (6.82 $\pm$ 0.01)$^{\dag}$ & (1.69 $\pm$ 0.01)  & (1.83 $\pm$ 0.01)&$\sim$1.12\\
  (EM register)& $\times$ 10$^{-4}$ &  $\times$ 10$^{-5}$ & $\times$ 10$^{-4}$ & $\times$ 10$^{-4}$&\\
\hline \\
EPER Par. CTI & (1.13 $\pm$ 0.20) & (5.40 $\pm$ 0.80) & (9.26 $\pm$ 0.70)  & (3.94 $\pm$ 0.45)& $\sim$5\\
& $\times$ 10$^{-4}$ &  $\times$ 10$^{-4}$ & $\times$ 10$^{-5}$ & $\times$ 10$^{-4}$&\\
\hline \\
EPER Serial CTI & (2.37 $\pm$ 0.70) & (2.06 $\pm$ 0.22) & (1.73 $\pm$ 0.21)  & (2.32 $\pm$ 0.38)& ... \\
& $\times$ 10$^{-4}$ &  $\times$ 10$^{-4}$ & $\times$ 10$^{-4}$ & $\times$ 10$^{-4}$&\\
\hline\\
\multicolumn{6}{c}{$^{\dag}$Device 1 did not have the EM register irradiated.  However, the CTI has been reduced}\\  
\multicolumn{6}{c}{by a factor of $\sim$1.7.  This may be because of a decrease in multiplication gain that was}\\
\multicolumn{6}{c}{observed for the device during post-irradiation testing.  The total amount of signal would thus}\\ 
\multicolumn{6}{c}{be smaller passing through the EM register and so less interaction with surface traps would occur.}\\
\multicolumn{6}{c}{Acronyms/abbreviations: Par. = parallel; CTI = Charge Transfer Inefficiency;}\\
\multicolumn{6}{c}{EPER = Extended Pixel Edge Response.}\\
\hline
\end{tabular}
\end{center}
\label{table:rad-results}
\end{table}

\subsubsection{Multiplication gain and read noise}

\textit{Multiplication gain.} The multiplication gain was measured
in the same way as system calibration, by using an $^{55}$Fe source
where the R$\phi$2HV voltage was varied. The gain was found for
each voltage by normalizing the system calibration gain to that
with a unity gain. The multiplication gain behaved as expected where we observed a measured gain with an exponential rise
with respect to the applied R$\phi$2HV clock. The gain profile
remained consistent for both devices before and after radiation
testing. We measured no difference (within error) of multiplication
gain for device 2, where we found values of 237 $\pm$ 8, prior
to and after radiation exposure. The serial register was irradiated
for this device, therefore, based on this result, it appears that this
level of fluence does not have an effect on the multiplication
gain for a room temperature, unbiased irradiation. However,
we note that if a causal flat-band voltage shift were to occur
as a result of such exposure, it could affect this result.

Interestingly, device 1 had a measured decrease in its multiplication
gain, even under a consistent R$\phi$2HV clock voltage.
This device was not subjected to any radiation exposure in
the serial register; therefore, this discrepancy cannot be attributed
to such effects. One plausible explanation is that device
aging might have caused degraded gain. This phenomenon
can occur due to the buildup of charge carriers under the
R$\phi$2HV electrode or the $\phi$DC electrode. Although e2v condition
their devices prior to release, the so-called ``burn-in phase'',
it is possible that aging could be responsible since the multiplication
gain for a given R$\phi$2HV voltage can decrease as a device
is continuously used. Aging will affect a device the most in
the first $\sim$10 h of life; however, some latent effects might still
be present. Finally, we note that a larger R$\phi$2HV voltage was
used for device 1 when compared to device 2. Even the smallest
increase in voltage in the high-gain register can cause significant
changes in amplification gain thereafter. Therefore, other transient
effects would be observed in device 1 before the same
effect might be observed in device 2. We are currently investigating
this further.\\

\textit{Read noise.} For this setup, the read noise of the CCD201 was
found in a similar way to that outlined in Sec. 5.2, where the 16 prescan elements (prior to the high-gain register) were used.
Although these pixels will still contain some dark current and
some CIC that has been amplified by the gain register from the
previous line readout, these components were removed by backclocking
the prescan elements before the line readout for a given
frame. A histogram of the signal as a function of DN can then be
used to find the read noise where a Gaussian curve is fitted and
the mean is calculated.

For devices 1 and 2, the readout noise was observed to
increase. As outlined in the previous subsection, since there
was a measured decrease in the multiplication gain of device
1, this might account for a subsequent increase in effective
read noise; however, since there was no measurable change
in device 2's multiplication gain, this explanation is not consistent
for both devices. Since device 2 was subjected to radiation
exposure in both the parallel and serial registers, dark current
generation in the serial register might have contributed to this
increase. However, measures were taken to minimize dark current
in device 2 by the back-clocking procedure described
above. Indeed, the stability of the EM process, and in turn
the multiplication gain and the read noise, all rely on careful
control of the electronics and device conditions throughout
an integration. Radiation damage can complicate this equilibrium.
As before, investigations are ongoing.

\subsubsection{Dark current (inverted mode operation)}

Long integration frames were obtained in the range 300 to
3600 s in order to measure dark current. CRs were removed
from the data as before, and parallel and serial CIC was removed
by using regions of the parallel overscan. The mean value over
a range of pixels in the image and store regions were then used
to calculate dark current in the parallel section. This was performed
pre- and postirradiation, and the results of this study
are shown in Figs. 15 and 16.

In each case, the measured dark current preirradiation was of
order 10$^{-5}$ e$^{-}$ pix$^{-1}$ sec$^{-1}$, where tests were performed at -108$^{\circ}$ C (165 K). Note that in the beam line, the sensor was at ambient
temperature and thus some of the subsequent damage from displacement
events was annealed out. The second phase of this
radiation study will irradiate under the same temperature at
all points of the study; this is outlined further later in this section.
Dark current was a factor of $\sim$1.8 higher for each device after
displacement damage had occurred. We highlight that device 1
was measured to have higher dark current than device 2. This
can be explained based on a classification of engineering grade
due to high dark current upon release from e2v. Therefore,
device 2 is likely more representative of what a science-grade
device would undergo in orbit. We note further that although
each device was measured in IMO, in this mode, the dark current
can be heavily dependent on the applied V$_{SS}$. As outlined previously,
we used standard CCD201 data sheet values for this
experiment. However, as noted in Fig. 10, the dark current in
this setup is consistent with the dark current trend from the
highly tuned EMN2 system that was used for sensor characterization
of CGI requirements.

\begin{figure}
   \centering
   \includegraphics[width=13cm]{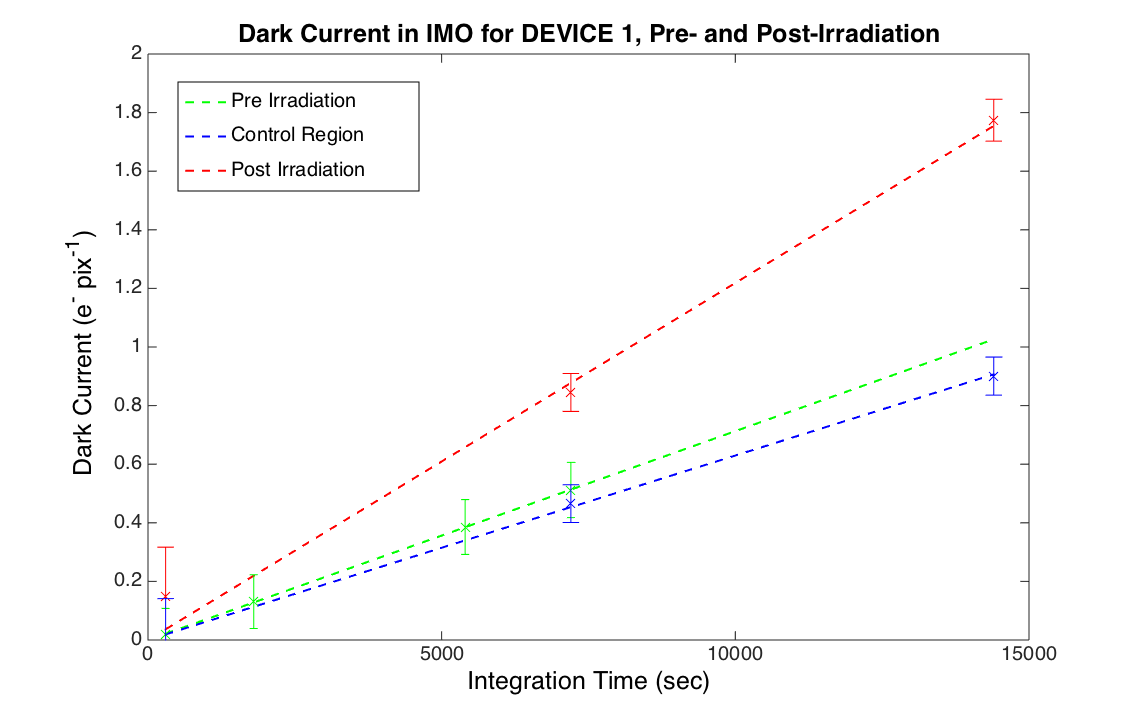}
\caption{Plot of dark current (e$^{-}$ pix$^{-1}$) as a function of integration time (s) for device 1 showing pre- and post-irradiation in addition to the control region.  We set the intercept to zero in order to assess clearly the trends for pre- and post-irradiation data.}\label{fig:IDK1}
\end{figure}

\subsubsection{Clock induced charge}\label{subsec:CIC-CEI}

\textit{Parallel clock induced charge.} Based on methods developed
by e2v, parallel CIC was measured by using a customized
sequencer. This is where a single line of signal is created in the
serial register by clocking a large number of rows. This signal is
then read out of the device, which is predominantly made up of
CIC. After this process, the device is subjected to a frame integration
time equal to the time taken to read out the number of
rows in the previous process. In this time, dark current will accumulate,
and these pixel rows are binned and also read out. As a
result, there are now alternate lines of dark current + CIC, and
dark current alone, for the same time interval. By calculating the
difference between these signals, the parallel CIC can be found.

A mean CIC value for each device was calculated by averaging
these rows and then dividing by the number of parallel
transfers. We obtained CIC measurements for different values of clock swing. The clock swing is the difference between the
high and low values in the image and store clocks. CIC is
heavily dependent on this parameter (based on higher local
electric field conditions for varying clock swing), but in this
work, we present pre- and postirradiation results for the lowest
CIC only. For both devices, the difference in CIC is very small,
as shown in Table 9. The expectation prior to the radiation test
was that CIC would not be affected by displacement damage
since the mechanism for the creation of CIC should not get
affected in this way. Trap sites are not responsible for the capture
and emission of holes during CIC generation; this happens
within the clocking process.We note that flat-band shifts should
be considered for device optimization, since any shifts of this
kind in flight would affect the derivation of the V$_{SS}$ value,
which can have a big effect on both dark current and CIC.
Indeed, the pinning potential, which refers to the V$_{SS}$, can
vary from device to device and thus different voltages may be
present. This might explain the discrepancy between the CIC
measured in devices 1 and 2, where CIC in device 1 was measured
to be higher.\\

\begin{figure}
   \centering
   \includegraphics[width=13cm]{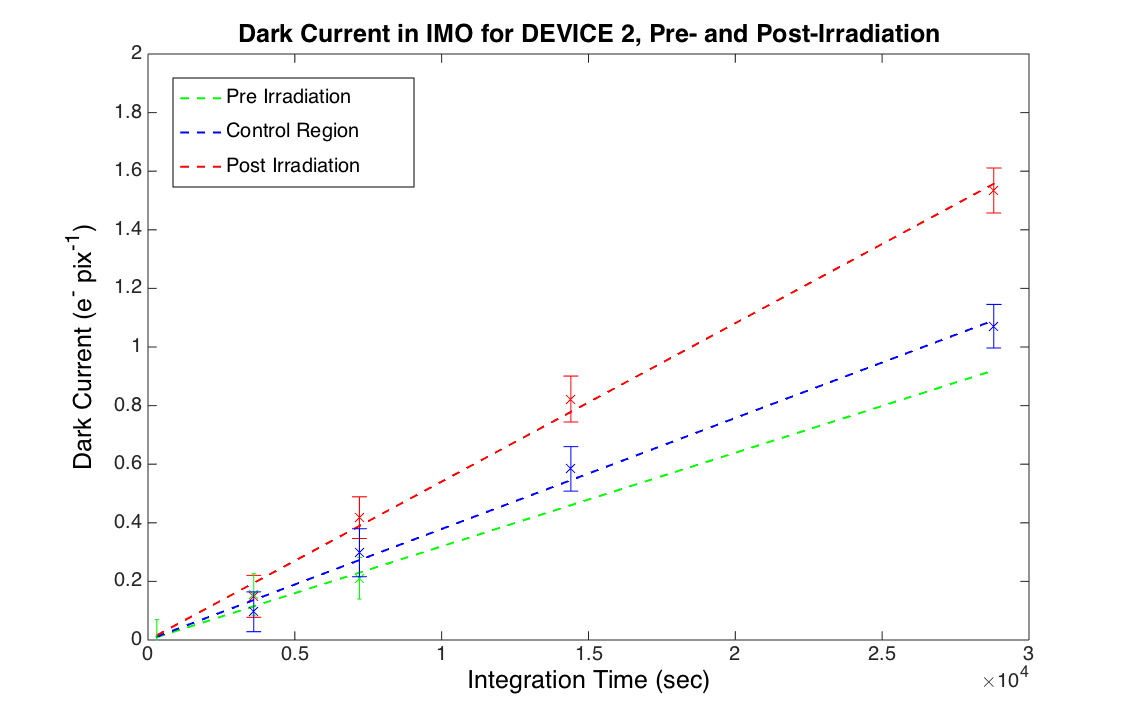}
\caption{Plot of dark current (e$^{-}$ pix$^{-1}$) as a function of integration time (s) for device 2 showing pre- and post-irradiation in addition to the control region.  As before, we set the intercept to zero in order to assess clearly the trends for pre- and post-irradiation data.}\label{fig:IDK2}
\end{figure}

\textit{Serial clock induced charge.} By considering the method
for calculating parallel CIC, the time taken to bin columns to
the output node was too long for the allowable CDS sampling
cycle of the control electronics; therefore serial CIC could not be
measured in the same way.We measured serial CIC as before, in
Sec. 5.4, by using the serial overscan in the images. This value
was then scaled by the system calibration measurements to calculate
CIC alone. However, we discovered some contribution
from other on-chip sources during this process (both pre- and
postirradiation) and do not consider the current analysis using
the test camera system to be providing representative values for
serial CIC alone. Investigation showed that this was a phenomenon
likely exacerbated by the camera system. We made sure to
confirm that the methods used to calculate other performance
figures take the noise into account and are unaffected.

\subsubsection{Charge transfer inefficiency}

\textit{$^{55}$Fe X-rays: high flux charge transfer inefficiency.} CTI
for high flux was measured using a $^{55}$Fe x-ray source by taking
50 images with integration times of 100 s at -108$^{\circ}$ C (165 K).
The x-ray density was $\sim$1 event per 200 pixels. Regions of the
sensor were assessed independently and the average value of all
single events in these regions was evaluated and plotted as a
function of pixel position in either row or column, depending
on whether parallel or serial CTI was being assessed. A fit
was applied to these data and the gradient of this fit was
used to calculate a value for CTI; the uncertainty in this derivation
was taken to be the standard deviation of single events
within a region. Both the conventional register (unity gain)
and the EM register were used to confirm that CTI was the
same in both cases (as expected). In Table 9, we show these
measurements.

As expected, devices 1 and 2 have shown increased CTI as
a result of displacement damage, the CTI measurements in the
control regions are consistent with the measured CTI in any
preirradiated region. We found no significant degradation of
the conventional register in device 1 (register was shielded),
whereas device 2 showed a measured decrease as a result of
the radiation dose, as expected, where we measure a factor of
$\sim$1.12 increased CTI in the EM register. We note that in order
to calculate the CTI in the EM register, some assumptions must
be made since the signal is changing as it moves through the
amplification process. These are (1) the CTI of each multiplication
element is the same and (2) the deferred tail, explained
below, accounts for most of the lost charge. Additionally, CTI in
the EM register will be dependent on the R$\phi$2HV and $\phi$DC
values, in addition to how much charge is present, the system
temperature and the gain applied.

\begin{figure}
   \centering
   \includegraphics[width=13cm]{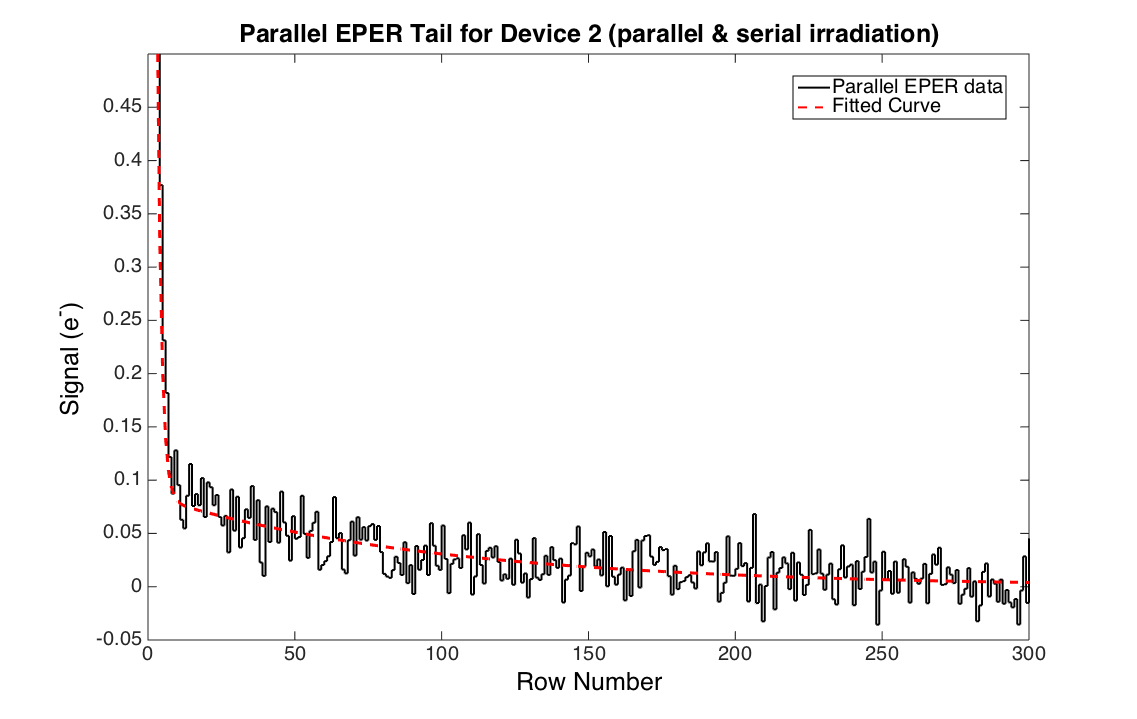}
\caption{Parallel EPER tail for device 2 (parallel and serial irradiation).  The first 300 deferred pixels are shown.  The fit consists of two exponential functions, where fast traps emitted in the first $\sim$10 pixels are fitted first then slower traps that are emitted some time later.  By integrating under the curve, the total charge emitted can be found, which is $\sim$8 e$^{-}$.}\label{fig:CTE1}
\end{figure}

To measure the deferred charge tail, images were once again
measured using a $^{55}$Fe x-ray source, with integration times of
100 s at -108$^{\circ}$ C (165 K). A histogram of the image was created
in order to identify single events, located at the Mn-K$\alpha$ peak (located by fitting a Gaussian). $\pm$2$\sigma$ limits were applied to the
fit, and events within this range were classified as single events.
All events were summed and averaged to estimate a mean value,
which helped to reduce the noise as a result of the ENF from
the multiplication process. The total deferred charge tail was
summed and used in the following equation\cite{daigle14} to calculate
the CTE in the EM register (we found the CTE via the following
equations and then converted to CTI as previously outlined,
where CTE $=$ 1 -- CTI):

\begin{equation}
CTE_{HV}=(1- E_{def})^{\frac{1}{N}},
\end{equation}
where $E_{def}$ is the deferred charge tail and $N$ is the total number
of multiplication elements in the CCD201, which is 604.
Crucially, much like CIC, the values for CTI obtained from
this experimental setup do not reflect those required for the
CGI. As previously stated, the appropriate considerations of
the R$\phi$2HV and $\phi$DC voltages, in addition to the operating
temperature, are all very important in minimizing the CTI of
a device. This study aimed to investigate degradation based
on displacement damage only.\\

\begin{figure}
   \centering
   \includegraphics[width=13cm]{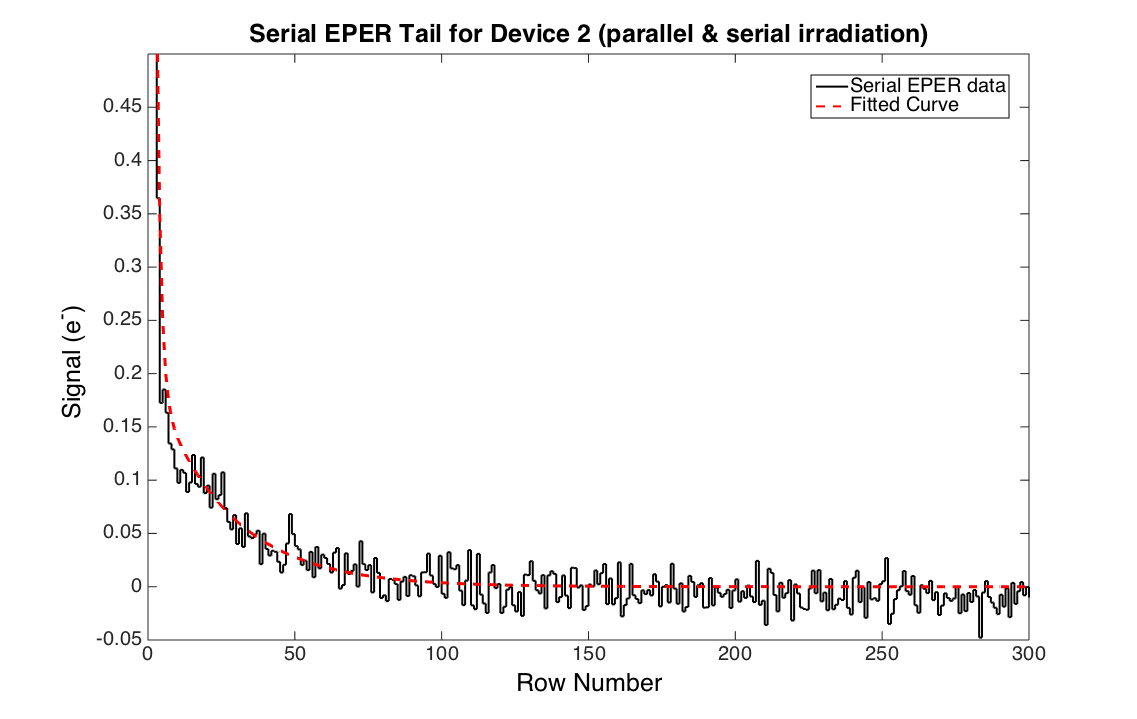}
\caption{Serial EPER tail for device 2 (parallel and serial irradiation), where as before, the first 300 pixels are shown.  Here the deferred charge is much more obvious than the parallel measurement in the first $\sim$100 pixels or so, likely due to probing different trap species at the different clock timings.}\label{fig:CTE2}
\end{figure}

\textit{Extended pixel edge response: low flux charge transfer inefficiency.} Low signal CTI, most applicable to the
WFIRST-CGI, was measured using the extended pixel edge
response (EPER) technique, see Janesick\cite{janesick01} for more details.
As before, we first find CTE and then convert to CTI. EPER
uses a flat field where the CTE is estimated by measuring
the amount of deferred charge found in the serial or parallel
overscan regions, defined as the line region (or ``extended
pixel'' region). The technique is based on the sharpness of
this edge and interpreting the results in terms of traps in the
detector causing degradation of charge transfer. If charge transfer,
as characterized by CTI is poor, large amount of charge from
the illumination get captured by traps and subsequently re-emitted
when the charge cloud is transferred. The drop in signal is known as a ``deferred charge tail'', which arises from this process.
If the charge transfer was perfect, there would be no
deferred charge, and a ``square wave'' edge would be present
in the EPER data. Since traps result in some of the charge
being deferred, instead of a perfectly square edge, there is a
tail comprising charge that went through trapping/release during
readout. CTE, in this respect, is calculated as follows:

\begin{equation}
CTE=1-\frac{S_{D}}{S_{LC}(e^{-}) \cdot N_{P}},
\end{equation}
where $S_{D}$ is the deferred charge, $S_{LC}$ is the charge in the last column read out, and $N_{P}$ is the number of pixel transfers. 

Importantly, we note that the signal collected with each pixel
passes through the conventional and EM registers before the
output, therefore the serial CTI represents a combination of
CTI from both. By only using the EPER method, it is not possible
to separate these components. The measured low-level CTI
is expected to be worse than the x-ray CTI method, and this
should be considered when interpreting the EPER results. As
before, CTE $=$ 1 -- CTI and we show results for CTE due to
the charge deferred in the trailing 300 pixels following the
flat field exposure of $\sim$8 e$^{-}$, see Figs. 17 and 18.

As shown in Table 9, parallel CTI for devices 1 and 2, preirradiation,
agrees within the error budget and are of order 10$^{-4}$.
We measure degradation of a factor of $\sim$5 postirradiation. We
cannot fully assess the effect of serial CTI, which appears to
degrade by a much smaller factor than parallel CTI. This
could mean that serial CTI in the context of this experiment
was dominated by the EM register as a result of higher operating
voltages and interaction with surface states based on the operating
conditions that were used for this particular measurement.

\subsection{Significance of Results for the WFIRST-CGI}

The results from the phase I radiation study are encouraging
when comparing the BOL and EOL detector performance relative to the requirements for the CGI. Based on current models
that predict planetary yield as function of integration time,
the most dominant sources of noise, or the performance parameters
of the device that we must pay particular attention to, are
dark current, CIC, and CTI. After annealing that followed the
phase I radiation testing, we can report that the CCD201 measured
dark current at operational temperatures of -108$^{\circ}$ C (IMO)
passes the CGI requirement both before and after a DDD
of 2.5 $\times$ 10$^{9}$ protons cm$^{-2}$, reflecting an EOL dose in L2 for
10 mm of tantalum shielding. Running at this temperature
may affect the CTI under high-gain conditions, and these effects
are discussed in the material that follows. Moreover, we can
report no measurable increase in CIC within measurement
error for pre- and postirradiation of both the parallel section
or the serial register for this unbiased, room temperature irradiation
under phase I. The results of CTI are still being considered,
and the operating conditions of the device will play a vital role
for optimization. In the following material, we outline the ``lessons
learned'' from the phase I study, as well as the phase II
study that commenced in June 2015.

\subsubsection{Lessons learned from phase I}\label{lessons}

The phase I experimental setup used standard CCD201 operating
voltages and bias conditions as provided by the e2v CCD201
specifications sheet. The phase II testing began in June 2015
and is outlined in Sec. 6.5.2. This study will greatly benefit
from ``lessons learned'' directly from phase I results, where
the information can be applied in a way that will optimize the
control electronics to maximize performance for all important
device parameters. Any recommendations for optimization
from phases I and II will be applied directly to the CGI strategy
in achieving the required S/N.

\begin{description}

  \item[$\bullet$ Multiplication Gain:] We found that the R$\phi$2HV and $\phi$DC
parameters can have a strong effect on the CTI of the gain
register of the device. Based on an operating temperature
of -108$^{\circ}$ C (165 K), running the R$\phi$2HV clock with lower
voltages can maintain better charge transfer; however, this is likely temperature dependent. Furthermore, a lower
setting in $\phi$DC can result in a higher gain for the same
R$\phi$2HV since the difference between these two parameters
will define the gain. If this is set too low, serial CTI
can increase.

 \item[$\bullet$ Dark Current:] Having prior knowledge of a specific device's
pinning potential, V$_{SPP}$, prior to irradiation would
be beneficial. Currently, the projected CGI operating
mode will be in IMO due to greater suppression of surface
dark current. In IMO, V$_{SS}$ is raised to a point where the
semiconductor–insulator interface is subjected to large
populations of holes that subsequently capture electrons,
thereby suppressing surface traps from emitting such
charge. We found that if V$_{SS}$ $>$ V$_{SPP}$, the parallel CIC
may be affected, where there will be a tradeoff between
dark current and CIC in this case. Measurements of dark
current and CIC as a function of V$_{SS}$ may elucidate this
potential optimization.
 
  \item[$\bullet$ Parallel \& Serial CIC:] Clock amplitudes and precision
timings in the controller are essential to minimizing the
contribution of CIC. Indeed, the CCCP controller is
highly tuned in this respect. As expected, for the electronics
used during phase I, we observed that operating at
lower clock amplitudes and faster speeds reduced the CIC
in both the parallel and serial sections. However, much
like dark current and CIC re: V$_{SS}$, there is a tradeoff with
CIC and CTI re: clock amplitudes and timing. Multilevel
clocking techniques have shown promising results in
minimizing CTI while reducing CIC.\cite{murray13} Importantly, by
combining multilevel clocking with a design modification
(as outlined in Secs. 5.5.4 and 5.5.6), one can increase
the parallel CIC to negligible levels with reduced CTI.
This work is ongoing and will be considered for the
WFIRST-CGI.

\item[$\bullet$ CTI:] These subsections have already outlined useful
optimizations for charge transfer. As noted previously, by adjusting many of these operating parameters, there is
likely a tradeoff between one performance criteria and
another, e.g., CIC/CTI or CIC/dark current. Once optimized
for flight at BOL, the primary concern for degradation
of charge transfer, characterized by CTI, will be the
capture and emission time constants for different trap species
within the buried channel, coupled with the device
operating temperature. Furthermore, different EMCCD
modes, such as photon-counting or analog mode, may be
affected in different ways, where one mode may be recommended
earlier in the mission under less damage, than
another. Such trap properties are vital when considering
clock speeds and other clocking parameters, in addition to
wave shapes and operating temperature. Optimization of
these parameters is ongoing at JPL where different design
recommendations will be put forward based on the
accrued damage from BOL and EOL.

\end{description}

\subsubsection{Phase II irradiation plan}\label{subsec:phaseII}

The phase II irradiation study took place at the Helios-3 beamline
in Harwell, United Kingdom, and was the first of its kind for
the CCD201 for the reasons outlined in the following material.
This new plan commenced on June 01, 2015, and consisted of: \\ \\\textbf{1)} A preirradiation characterization of two science-grade CCD201s.\\ \\ \textbf{2)} Four separate irradiation and characterizations at the following proton fluences: 1.0 $\times$ 10$^{9}$, 2.5 $\times$ 10$^{9}$, 5.0 $\times$ 10$^{9}$ and 7.5 $\times$ 10$^{9}$ protons cm$^{-2}$ that reflect the full range of radiation doses that a sensor would undergo during a 6-year flight in L2, by considering both tantalum and aluminum shielding of varied thicknesses (see Fig. 6). \\ \\  \textbf{3)} During the irradiation process at Harwell, the sensor was maintained at a temperature of -108$^{\circ}$ C (165 K) for all fluences and was kept at this temperature during characterization in order to avoid annealing the effects of damage. A vacuum chamber has been designed to operate in the Harwell beamline for this reason.\cite{gow15} \\ \\ \textbf{4)} The CCD was powered on during irradiation in order to investigate a positive flat-band shift as is expected in flight.\\ \\ \textbf{5)} Post-irradiation characterization of each sensor.  \\

Subsequent analysis of the radiation-damaged sensors at JPL
will further reveal the effects of annealing on the device following
characterization at Harwell. Additionally, thermal cycling
testing will also be carried out.

By assessing the sensor's radiation damage at a consistent
cryogenic temperature, the density of defects and populations
of the various trap species are expected to be different when
compared to a room temperature irradiation, subsequently
impacting dark current and CTI. We believe that this study,
coupled with what was learned from phase I, will be an accurate
representation of what an EMCCD will undergo in orbit by
considering the effects of displacement damage. The operating
modes mirrored the phase I setup, with the exception of some of the modifications that were outlined in Sec. 6.5.1.
For example:

\begin{itemize}

  \item The read noise was measured by using the CCD201's 16
prescan elements that are located adjacent to the high gain
output, see Fig. 3. In this test, the high voltage clock was
powered down as opposed to back-clocking the register,
resulting in no signal in this region and a clean measure of
the read noise from the output amplifier.

  \item The multiplication gain was measured only when it has
reached a stable condition, by using a customized
sequencer that has been developed at the CEI. By passing
large amounts of signal through the EM register prior to
full testing, the multiplication gain can be accurately
measured since this prevents any transient effects that
may impact the R$\phi$2HV clock at the beginning of a read
out sequence, e.g., thermal stability of electronics.
  
\end{itemize}

The primary proton beam energy at Harwell is 7.5 MeV, with
a flux of $\sim$ 1.0 $\times$ 10$^{7}$ protons cm$^{-2}$ sec$^{-1}$. Much like PSI, the
dosimetry is expected to be $<$10\% across the full sensor area.
Because the primary beam is of energy $<$13.5 MeV, the NIEL
equations [refer to Eqs. (6)--(9)] used previously will differ:

\begin{equation}\label{eq:NIEL_new}
\mbox{10 MeV NIEL Function}=\frac{8}{E_{p}^{0.9}},
\end{equation}
where $E_{p}$ is the primary beam energy (MeV), yielding a total
fluence of 1.92 $\times$ 10$^{9}$ protons cm$^{-2}$ sec$^{-1}$. Furthermore, it is possible
that some equipment may become activated by the primary
beam. Since this equipment may be radioactive as a result, postirradiation
only commenced once the surrounding conditions
were safe for personnel and to avoid additional signal in the
images. Prior to irradiation, an engineering grade device has
undergone activation tests to assess the stability of defects in
the device (e.g., dark current), so that a plan is in place postirradiation
to assess the effects of annealing if necessary.\\

\textit{Thermal cycling at JPL.} After initial postcharacterization
has been completed at the CEI, the irradiated science-grade
CCD201s will be shipped to JPL where they will undergo multiple
thermal cycles over the projected mission survival temperature
range. Furthermore, each device will undergo a thermal
soak at $>$50$^{\circ}$ C and at -120$^{\circ}$ C for 24 h for survival testing.
Automated testing will be designed and implemented in
order to do so. Full device characterization will be carried
out post-thermal soak, paying particular attention to the effects
of annealing.

\section{CONCLUSIONS}
\label{sec:discussion}  % \label{} allows reference to this section

Here, we present performance characterization of the CCD201-
20 EMCCD that has been baselined for the WFIRST-CGI.
This characterization has been carried out for BOL conditions,
as well as conditions that reflect the space environment of
the CGI at the EOL of a 6-year flight in L2, specifically,
displacement damage due to protons assuming a 10-mm tantalum
shield. In the case of BOL characterization to meet the
CGI detector requirements, we use the N{\"u}v{\"u} Cameras EMN2
camera system with the CCCP controller, which uses highly
tuned clocking to achieve subelectron read noise, dark current
of order 5 $\times$ 10$^{-4}$ e$^{-}$ pix $^{-1}$ sec$^{-1}$ at -85$^{\circ}$ C in IMO, and CIC
of order 10$^{-3}$ e$^{-}$ pix $^{-1}$ fr$^{-1}$. The measurement of initial radiation-induced performance degradation was carried out at
the PSI in Switzerland using generic drive electronics from
XCAM Ltd., and parameters, such as multiplication gain,
dark current, CIC and CTI, were assessed, as well as their
impact on the CGI application. We report degradation of
dark current of a factor of $\sim$1.8, which still meets CGI requirements
for a temperature of -108$^{\circ}$ C at EOL. Similarly, no degradation
of CIC was observed following room temperature
irradiation, and we are still assessing the effect of charge transfer
degradation of a factor of $\sim$1.12 in the EM register and $\sim$5 in
the parallel section. The phase I study has provided important
information on the required operating parameters in order to
minimize dark current, CIC and CTI, as well as highlighting
trade-offs between voltage/bias conditions and precision timing
of an EMCCD controller, in order to provide the best overall
S∕N performance for the WFIRST-CGI. A second phase
of the radiation study commenced in June 2015 and was completed
in August 2015, which was the first of its kind for a
CCD201-20. The device was irradiated over the full range of
proton fluences reflecting BOL to EOL for L2 that reflect
realistic conditions of the WFIRST-CGI EMCCD sensors, while
kept at cryogenic temperatures where the device was under
power to measure flat-band shifts as expected in flight. These
results will be released following subsequent analysis.

%%%%%%%%%%%%%%%%%%%%%%%%%%%%%%%%%%%%%%%%%%%%%%%%%%%%%%%%%%%%%
\acknowledgments     %>>>> equivalent to \section*{ACKNOWLEDGMENTS}       
 
Beginning-of-life device performance characterization was carried
out at the Jet Propulsion Laboratory, California Institute of
Technology, under a contract with the National Aeronautics
and Space Administration. The irradiation component of this
work was carried out at the Proton Irradiation Facility (PIF)
at PSI, Switzerland, under the guidance of the Jet Propulsion
Laboratory and the Center for Electronic Imaging. We would
like to extend special thanks to the staff of the beamline facility
at PSI, Switzerland. The authors acknowledge the support of
N{\"u}v{\"u} Cameras regarding the EMN2 system, as well as helpful
conversations with Anders Petersen, Bernard Rauscher, and
Navtej Singh, and with device experts at e2v Technologies,
Chelmsford, UK. Finally, we thank the journal editor in addition
to the referees for their careful reading of our work and for their
helpful input on how to improve this paper.

%%%%%%%%%%%%%%%%%%%%%%%%%%%%%%%%%%%%%%%%%%%%%%%%%%%%%%%%%%%%%
%%%%% References %%%%%

\end{document}